\documentstyle[aaspp4,flushrt]{article} \input epsf

\begin{document}
\newcommand\approxgt{\mbox{$^{>}\hspace{-0.24cm}_{\sim}$}}
\newcommand\approxlt{\mbox{$^{<}\hspace{-0.24cm}_{\sim}$}}

\newcommand{\vertsp}{\vphantom{\displaystyle{\dot a \over a}}}
\newcommand{\se}{{(0)}}
\newcommand{\ve}{{(1)}}
\newcommand{\te}{{(2)}}
\newcommand{\nnu}{\nu}
\newcommand{\Spy}[3]{\, {}_{#1}^{\vphantom{#3}} Y_{#2}^{#3}}
\newcommand{\Gm}[3]{\, {}_{#1}^{\vphantom{#3}} G_{#2}^{#3}}
\newcommand{\Spin}[4]{\, {}_{#2}^{\vphantom{#4}} {#1}_{#3}^{#4}}
\newcommand{\scpot}{{\cal V}}
\newcommand{\tl}{\tilde}
\newcommand{\bm}{\boldmath}
\def\bi#1{\hbox{\boldmath{$#1$}}}

% redefine ell -> l for PRD
\renewcommand{\ell}{l}
\renewcommand{\topfraction}{1.0}
\renewcommand{\bottomfraction}{1.0}
\renewcommand{\textfraction}{0.00}
\renewcommand{\dbltopfraction}{1.0}

\newcommand{\beq}{\begin{equation}}
\newcommand{\eeq}{\end{equation}}
\newcommand{\beqa}{\begin{eqnarray}}
\newcommand{\eeqa}{\end{eqnarray}}

\newcommand{\lexp}{\mathop{\langle}}
\newcommand{\rexp}{\mathop{\rangle}}
\newcommand{\rexpc}{\mathop{\rangle_c}}

\def\d{\delta}
\def\te{\theta}
\def\ds{\delta_s}
\def\dt{\tilde \delta}
\def\dD{\delta_{\rm D}}
\def\del{\nabla}
\def\knl{k_{n\ell}}
\def\pl{{\mathsf P}}
\def\nb{\bar n}

\font\BF=cmmib10
\font\BFs=cmmib10 scaled 833
%\font\BF=cmmib10 scaled 1200
%\font\BFs=cmmib10
\def\k{{\hbox{\BF k}}}
\def\x{{\hbox{\BF x}}}
\def\r{{\hbox{\BF r}}}
\def\s{{\hbox{\BF s}}}
\def\ks{{\hbox{\BFs k}}}
\def\xs{{\hbox{\BFs x}}}
\def\q{{\hbox{\BF q}}}
\def\v{{\hbox{\BF v}}}
\def\u{{\hbox{$u_z$}}}
\def\tk{\hat k}
\def\tvk{{\hat{\k}}}
\def\ttheta{\hat \theta}
\def\tphi{\hat \varphi}

\def\la{\mathrel{\mathpalette\fun <}}
\def\ga{\mathrel{\mathpalette\fun >}}
\def\fun#1#2{\lower3.6pt\vbox{\baselineskip0pt\lineskip.9pt
\ialign{$\mathsurround=0pt#1\hfill##\hfil$\crcr#2\crcr\sim\crcr}}}

%\draft
\hfill{\small CITA-98-62, FERMILAB-Pub-99/003-A}

\title{Power Spectrum Correlations Induced by Non-Linear Clustering}

\author{Rom\'{a}n Scoccimarro$^{1,\dag}$, Matias Zaldarriaga$^{2,\ddag}$ \&
Lam Hui$^{3,\P}$}

\vskip 1pc

\affil{${}^\dag${\sf scoccima@cita.utoronto.ca};\ \ ${}^\ddag${\sf
matiasz@ias.edu};\ \ ${}^\P${\sf lhui@fnal.gov}}

\affil{${}^1$CITA, McLennan Physical Labs, 60 St George Street,
Toronto ON M5S 3H8, Canada}

\affil{${}^2$Institute for Advanced Study, School of Natural Sciences,
Olden Lane, Princeton, NJ 08540}

\affil{${}^3$NASA/Fermilab Astrophysics Center, Fermi National
Accelerator Laboratory, Batavia, IL  60510}

\begin{abstract}

Gravitational clustering is an intrinsically non-linear process that
generates significant non-Gaussian signatures in the density field.
We consider how these affect power spectrum determinations from galaxy
and weak-lensing surveys.  Non-Gaussian effects not only increase the
individual error bars compared to the Gaussian case but, most
importantly, lead to non-trivial cross-correlations between different
band-powers, correlating small-scale band-powers both among themselves
and with those at large scales.  We calculate the power-spectrum
covariance matrix in non-linear perturbation theory (weakly non-linear
regime), in the hierarchical model (strongly non-linear regime), and
from numerical simulations in real and redshift space.  In particular,
we show that the hierarchical ansatz cannot be strictly valid for the
configurations of the trispectrum involved in the calculation of the
power-spectrum covariance matrix.

We discuss the impact of these results on parameter estimation from
power-spectrum measurements and their dependence on the size of the
survey and the choice of band-powers. We show that the non-Gaussian
terms in the covariance matrix become dominant for scales smaller than
the non-linear scale $k_{nl} \sim 0.2$ h/Mpc, depending somewhat on
power normalization. Furthermore, we find that cross-correlations
mostly deteriorate the determination of the amplitude of a rescaled
power spectrum, whereas its shape is less affected.  In weak lensing
surveys the projection tends to reduce the importance of non-Gaussian
effects. Even so, for background galaxies at redshift $z \sim 1$, the
non-Gaussian contribution rises significantly around $l\sim 1000$, and
could become comparable to the Gaussian terms depending upon the power
spectrum normalization and cosmology. The projection has another
interesting effect: the ratio between non-Gaussian and Gaussian
contributions saturates and can even decrease at small enough angular
scales if the power spectrum of the 3D field falls faster than
$k^{-2}$.

\end{abstract}

\keywords{large-scale structure of universe; methods: numerical;
methods: statistical}

\clearpage

\section{Introduction}
\label{intro}

Discussions in the literature on the measurement of the power spectrum
$P(k)$ (e.g. \cite{FKP94}), and the estimation of cosmological
parameters from it (e.g.  \cite{Tegmark97b,EHT98,HuTe98}), have
largely focused on the case of Gaussian random data.  While this is a
reasonable assumption in the case of the cosmic microwave background
for Gaussian initial conditions (e.g.
\cite{Tegmark97,BJK97,Seljak97}), it is probably not a good
approximation for galaxy and weak-lensing surveys except on
sufficiently large scales (e.g.
\cite{FKP94,VS96,Hamilton97,CSS97,Seljak97b,THSVS97,DHJ97}).  This is
because of the non-Gaussianity inevitably induced by gravitational
clustering, which leads to increased error bars in individual
band-power estimates and introduces correlations between them.  In
this paper, we quantify the size of these effects and the scales at
which they become important.

Understanding the statistical properties of power-spectrum estimators
beyond the Gaussian approximation requires at least a calculation of
the power-spectrum covariance matrix, which involves the four-point
function of the density field in Fourier space, the {\em
trispectrum}. In order to do this, we use both analytic and numerical
techniques.  At weakly non-linear scales, non-linear perturbation
theory (PT) can be used to understand quantitatively how the
non-Gaussian effects are generated through mode-coupling. We provide a
calculation of the relevant configurations of the trispectrum, and use
them to obtain the covariance matrix of band-powers.

Non-Gaussian effects are most significant at non-linear scales, where
PT breaks down. To investigate this regime, we resort to numerical
simulations, which we use to evaluate the power-spectrum covariance
matrix by measuring the power spectrum in several realizations. At
small scales, where virialization is reached, stable clustering
suggests a simple behavior of higher-order correlation functions,
known as the hierarchical ansatz. We use these arguments to understand
the power-spectrum covariance matrix in the non-linear regime, which
in turn allows us to extend in a simple way our results to the
projected density field. 

The outline of the paper is as follows.  In \S \ref{def}, we introduce
the notations and basic expressions for the covariance matrix of the
3D (three-dimensional) power-spectrum estimates.  Here, as in the rest
of the paper, we focus on the sample-variance dominated regime.  In
other words, shot-noise is ignored - the emphasis here is on the
effect of non-Gaussianity induced by gravitational
clustering. Shot-noise will be included in a separate treatment. We
present estimates of the power-spectrum covariance matrix using PT,
numerical simulations and the hierarchical model in \S \ref{PT},
\ref{Nbody} and \ref{hier}, respectively. In particular, we show in
the latter that the hierarchical ansatz does not provide a good
description of the covariance matrix in the non-linear regime, because
there are always contributions coming from large scales (\S
\ref{4ptf}). Instead, we find that the covariance matrix behaves as a
``generalized'' hierarchical model, with configuration dependence that
resembles that in PT (Figs. 9-11). On the other hand, for trispectrum
configurations fully in the non-linear regime, we verify the validity
of the hierarchical ansatz and provide some new results concerning the
importance of different contributions (Fig.8).

The impact of non-Gaussianity on parameter determination is
illustrated in \S \ref{parameter} with a simple example, in which the
amplitude of the power spectrum is the only parameter of interest. We
discuss the impact on the determination of the power spectrum shape as
well. In all of the discussions above, it is implicitly assumed one
can measure the mass power spectrum directly.  This is of course
complicated by biasing in the case of galaxy surveys.  In fact, as we
shall see, on scales where non-Gaussianity is non-negligible,
galaxy-biasing is also likely to be non-trivial.  Weak gravitational
lensing (e.g.
\cite{blandford91,mir91,Kaiser92,BVWM97,JaSe97,Kaiser98,Seljak97b,VWBM98})
promises to be a bias-free way to measure the mass power spectrum, but
as we will show in \S \ref{lensing}, weak-lensing surveys in the near
future will likely cover small angular scales and thus receive
significant contributions from non-linear fluctuations to the expected
signals.  In \S \ref{lensing}, we derive the appropriate expression
for the covariance matrix of the projected mass power spectrum, and
show that the projection reduces the non-Gaussian effects of the
matter distribution, but non-negligible residuals remain.  In fact,
for low matter density models where the cluster normalization implies
larger {\em rms} density fluctuations, the non-Gaussian contribution
to the covariance can dominate over the Gaussian value
(\cite{VWBM98}).  Finally, we conclude in \S \ref{discuss} with a
discussion.

\section{Definitions}
\label{def}

We shall be interested in the Fourier modes of the overdensity
$\d(\bi x)$ given by,
\begin{eqnarray}
\d(\bi k)=\int \frac{d^3{\x}}{(2\pi)^3}\ e^{-i\bi k \cdot \bi x} \
\d(\bi x) \, .
\label{deltak}
\end{eqnarray}
Their statistical properties are described by the various connected
moments,  \begin{eqnarray}
\langle \d(\k_1)\rangle_c&=&0 \nonumber \\ \langle
\d(\k_1)\d(\k_2)\rangle_c&=&
\dD(\k_{12})\ P(\k_1)\nonumber \\
\langle \d(\k_1)\d(\k_2) \d(\k_3)\rangle_c&=& \dD(\k_{123}) \
B(\k_1,\k_2,\k_3) \nonumber \\
\langle \d(\k_1)\d(\k_2) \d(\k_3)\d(\k_4)\rangle_c &=&
\dD(\k_{1234})\ T(\k_1,\k_2,\k_3,\k_4)\nonumber \\
\langle \d(\k_1)\d(\k_2) \cdots \d(\k_n)\rangle_c &=&
\dD(\k_{12\cdots n})\ T_n(\k_1,\k_2,\cdots,\k_n),
\label{npoint}
\end{eqnarray}
where $\k_{i...j} \equiv \k_i+\cdots +\k_j$, with $\dD(x)$ denoting
the Dirac delta distribution. In equation (\ref{npoint}), $P(k)$,
$B(\k_1,\k_2,\k_3)$ and $T(\k_1,\k_2,\k_3,\k_4)$ denote the power
spectrum, bispectrum and trispectrum, respectively.  For the purpose
of this paper we will need up to the fourth connected moment.

Let's suppose we are given a survey with volume $V$, from which $\d(\bi
k)$ can be constructed. We divide the Fourier space into shells (or
bands) of width $\d k$ centered on $k_i=i\times \d k, i=1,2,...$, and
then average the variance of the $\k$ modes within each shell to
obtain the following estimate of the power spectrum, \

\begin{eqnarray}
{\hat P}(k_i)&=& V_f \int_{k_i} {d^3\k\over V_s(k_i)} \ \d(\k)\d(-\k)
\label{estim}
\end{eqnarray}
where the integration extends over modes within the shell centered at
$k_i$, $V_s(k_i)=4\pi k_i^2 \d k$ is the volume of the shell, and
$V_f$ is the volume of the fundamental cell in $k$ space,
$V_f=(2\pi)^3/V$. For the rest of this paper, we will also use $P_i$
interchangeably with $P(k_i)$ wherever confusion will not arise.  It
is straightforward to calculate the covariance matrix of the power
spectrum estimators. Combining equations (\ref{npoint}) and
(\ref{estim}), we get

\begin{eqnarray}
\label{cij}
{C}_{ij}&\equiv& \langle {\hat P}(k_i){\hat P}(k_j) \rangle - \langle
{\hat P}(k_i)\rangle \langle{\hat P}(k_j) \rangle =V_f\ [{2
P_i^2\over V_s(k_i)} \d_{ij} + {\bar T}(k_i,k_j)]
\end{eqnarray}
where $\d_{ij}$ is the Kronecker delta and $\bar T$ is the bin-averaged
trispectrum

\begin{eqnarray}
{\bar T}(k_i,k_j)&=&\int_{k_i} {d^3{\k_1}\over V_s(k_i)}\ \int_{k_j}
{d^3{\k_2}\over V_s(k_j)}\ T(\k_1,-\k_1,\k_2,-\k_2). \label{cova1}
\end{eqnarray}
The first term in equation (\ref{cij}) is the Gaussian
contribution. In the Gaussian limit, each Fourier mode is an
independent Gaussian random variable.  The power estimates of
different bands are therefore uncorrelated, and the covariance is
simply given by $2/N_{k_i}$ where $N_{k_i}$ is the number of
independent Gaussian variables, i.e. $N_{k_i} = V_s(k_i)/V_f$. The
second term in equation (\ref{cij}) arises because of non-Gaussianity,
which generally introduces correlations between different Fourier
modes, and hence it is not diagonal in general.

Both terms in the covariance matrix in equation (\ref{cij}) are
proportional to $V_f$, or inversely proportional to $V$, the volume of
the survey. But the Gaussian and non-Gaussian contributions scale in a
different way with $V_s$: while the Gaussian contribution decreases
with the size of the shell, the non-Gaussian term remains
constant. Therefore, when the covariance matrix is dominated by the
non-Gaussian contribution the only way to reduce the variance of the
power spectrum is to increase the volume of the survey instead of
averaging over more Fourier modes.

It is worth emphasizing here that we have ignored the effect of the
survey window in the above expressions. We have implicitly assumed the
modes of interest have wavelengths much smaller than the size of the
survey, which is likely to be a good approximation as we are
interested in the non-Gaussianity induced by gravity on small scales
(how small is small is a question we will address). Note also that
while $\delta(\k)$ is not restricted to be the mass overdensity in the
above expressions, we will assume so in the rest of the paper.
Galaxy surveys of course only probe directly the galaxy rather than
the mass overdensity. For the most part, we ignore galaxy biasing in
this paper.  We will come back to this issue in the final section.

\section{Covariance Matrix of the Band-Power Estimates in 3D}
\label{results}

In structure formation scenarios where the initial conditions are
Gaussian, the power-spectrum covariance matrix is expected to be
diagonal on sufficiently large scales, with amplitude given by the
first term in equation (\ref{cij}). Gravitational clustering, however,
inevitably generates a non-vanishing trispectrum through non-linear
mode-coupling. It is thus {\em expected} that at small enough scales, the
non-Gaussian contribution to the power-spectrum covariance matrix
(second term in equation [\ref{cij}]) will dominate over the Gaussian
term.

To demonstrate this, we use non-linear perturbation theory (PT) to
calculate at what scales significant non-Gaussian contributions
first appear.  We then use numerical simulations to measure their
effects on smaller, non-linear scales, where PT breaks down.  In
the highly non-linear regime we study the validity of the
hierarchical model for the trispectrum, and calibrate it using
N-body simulations and Hyperextended PT (\cite{ScFr98}).

\subsection{The Power-Spectrum Covariance Matrix in PT} \label{PT}

In tree-level (leading-order) PT, the trispectrum is given by (see
e.g. \cite{Fry84}):

\beqa
T(\k_1,\k_2,\k_3,\k_4) &=& 4[ F_2(\k_{12},-\k_1) \ F_2(\k_{12},\k_3)
\ P_1 P_{12} P_3 + {\rm cyc.}] + 6[ F_3(\k_1,\k_2,\k_3) \ P_1 P_2 P_3
+ {\rm cyc.} ],
\label{4pt}
\eeqa

\noindent where $P_i \equiv P(k_i)$, with $\k_{12} =\k_1+\k_2$ as
defined before, ``cyc.'' denotes cyclic permutation of the arguments
(12 terms in total in the first contribution, and 4 terms in the
second contribution), and the kernels $F_n$ are obtained from solving
the equations of motion of gravitational instability to $n^{\rm
th}$ order in PT (\cite{GGRW86}).
Equation (\ref{cova1}) tells us that only a particular class of
trispectrum configurations (parallelogram) contributes to the
non-Gaussian covariance of the band-power estimates. Substituting the
above expression into equation (\ref{cova1}), we have
\begin{eqnarray}
{\bar T}(k_i,k_j)&=&\int_{k_i} {d^3{\k_1}\over V_s(k_i)}\ \int_{k_j}
{d^3{\k_2}\over V_s(k_j)}\ \Big[ 12 F_3(\k_1,-\k_1,\k_2) P_1^2 P_2 +
8 F_2^2(\k_1-\k_2,\k_2) P(|\k_1-\k_2|) P_2^2 \nonumber \\ &+& 16
F_2(\k_1-\k_2,\k_2) F_2(\k_2-\k_1,\k_1) P_1 P_2 P(|\k_1-\k_2|) + (\k_1
\leftrightarrow \k_2) \Big]
\label{cova2}
\end{eqnarray}

Using the facts that $[F_2(\k_1-\k_2,\k_2)]_{k_1=k_2} = (3+10x)/28$,
where $x \equiv {\hat k}_1 \cdot {\hat k}_2$, and that the
angular average $\lexp F_3(\k_1,-\k_1,\k_2) \rexp_{k_1=k_2} =
-11/378$, plus the approximate angular-averaged result

\beq
\int_{-1}^{1} \frac{dx}{2}\ \Big( \frac{3+10 x}{28} \Big)^2
P(\sqrt{2(1-x)}k) \approx \frac{30}{784}\ P(k),
\eeq
a simple approximate expression for the diagonal components
of the band-averaged trispectrum follows

\beq
\label{PTdiag}
 {\bar T}(k_i,k_i) \approx \frac{232}{441}\ P_i^3 \, ,
\eeq
where we have further assumed that the power spectrum is approximately
constant within the shell. Note that the non-Gaussian terms scale as
the power spectrum cubed, therefore, to be consistent within PT,
one-loop corrections to the power spectrum in the diagonal Gaussian
term must be included.

\begin{figure*}
\begin{center}
\leavevmode
\epsfxsize=4.0in \epsfbox{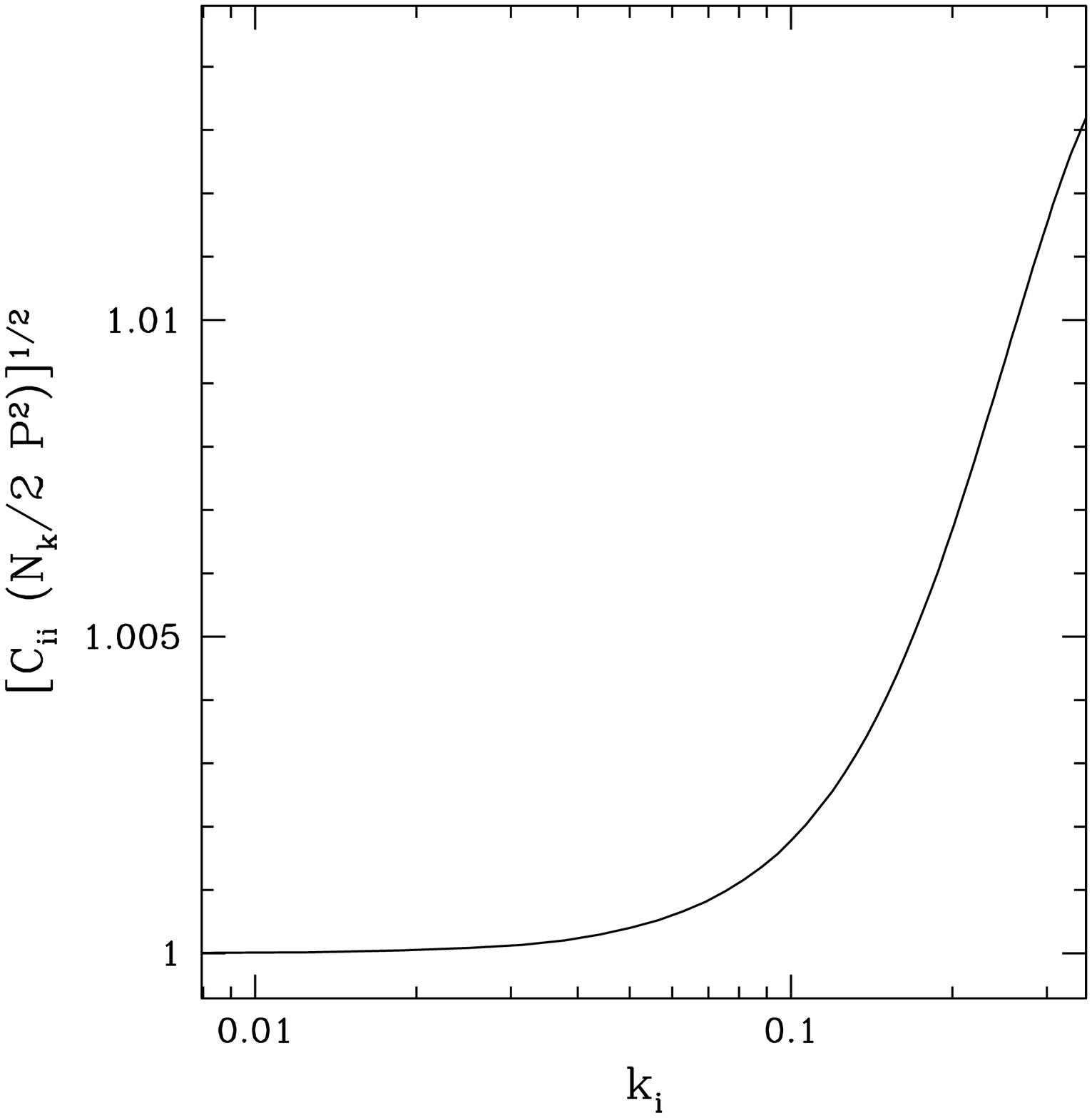}
\end{center}
\caption{Diagonal elements of the power spectrum covariance matrix
normalized by the Gaussian variance for SCDM with $\sigma_8=0.60$
($z=0$).  The centers of the shells are given by $k_i=(2\pi i)$ h/Gpc,
and width $\d k=2\pi$ h/Gpc.  For this model the non-linear scale is
$k_{nl}=0.33$ h/Mpc.}
\label{PT1}
\end{figure*}

Figure \ref{PT1} shows the diagonal elements of the covariance matrix,
divided by the Gaussian contribution, calculated using perturbation
theory for a survey with $V = 1 {\rm (Gpc/h)^3}$ .  The cosmological
model we assume throughout the paper is standard cold dark matter
(SCDM) with $\sigma_8=0.60$.  The increase in the diagonal variance
relative to the Gaussian variance is very small, less than $1\%$ for
$k<0.2$, which in part reflects the small bin-size considered, $\d k
=2\pi$~h/Gpc.  The non-linear scale, where $4 \pi k^3 P(k) = 1$, is
given by $k_{nl} = 0.33$.  It is important to emphasize that
figure \ref{PT1} depends on the binning adopted.  In other words, the
relative importance of the Gaussian and non-Gaussian variances for
each band is dependent upon the size of the band.

In figure \ref{PT2} we show some of the off-diagonal elements, in
terms of the cross-correlation coefficient $r_{ij}\equiv C_{ij}/
\sqrt{C_{ii}C_{jj}}$.  Each curve in figure \ref{PT2} corresponds to
$r_{ij}$ as a function of $k_i$ for a fixed $k_j$.  The corresponding
$k_j$ can be inferred from where $r_{jj}=1$.  By definition each
$r_{ij}$ is independent of the volume of the survey but it does depend
on the $k$-space binning.  In this case, the binning has been chosen
to be constant with the smallest possible value, $\d k=2\pi$ h/Gpc,
which maximizes the Gaussian contribution.  The correlation
coefficients $r_{ij}$, for $i \ne j$, are therefore quite small.  A
logarithmic binning would change the appearance of the figures.  It is
important to emphasize that although the correlation coefficients are
small, there are also many of them.  In fact because the size of the
correlation coefficients depends on the choice of band-powers, they do
not have a direct physical meaning.  In the limit we are considering
in this section, where the diagonal terms of the covariance are
dominated by the Gaussian contributions we have in fact $r_{ij}\propto
\delta k$.  In \S \ref{parameter} we will address the importance of
the non-Gaussian terms in a way independent of the binning procedure.

In order to understand the origin of power spectrum
cross-correlations, let us consider the integrand in
equation~(\ref{cova2}), $T(\k_1,-\k_1,\k_2,-\k_2) \equiv
T_4(\k_1,\k_2) $, and normalize it according to

\beq \beta_{12}(\k_1,\k_2) \equiv T_4(\k_1,\k_2)\
\sqrt{\frac{1}{2 P_1^2}\frac{1}{2P_2^2}
\frac{V_s(k_1)V_s(k_2)}{\Delta(k_1)\Delta(k_2)}}= \frac{1}{2}
\frac{T_4(\k_1,\k_2)}{(P_1 P_2)^{3/2}} \sqrt{ \frac{\d k_1 \d k_2}{k_1
k_2}},
\label{beta}
\eeq
where $\Delta(k) \equiv 4 \pi k^3 P(k)$. We then obtain

\beq C_{ij}= V_f\ \frac{2 P_i P_j}{\sqrt{V_s(k_i) V_s(k_j)}}\ [\d_{ij} +
{\bar \beta}_{ij} \sqrt{\Delta(k_i) \Delta(k_j)} ],
\eeq

\noindent with ${\bar \beta}_{ij}$ the bin-averaged version of
$\beta_{12}$.  Figure~\ref{corre} shows the coefficient $\beta_{12}$
as a function of configuration angle $\theta$ between $\k_1$ and
$\k_2$ for different scales, assuming constant bin size as a function
of scale $\d k_1=\d k_2=2\pi/100$~h/Mpc.  This figure illustrates how
modes get correlated, the maximum amplitude of correlation results for
co-linear configurations ($\theta=0$), and the least amount of
correlation corresponds to perpendicular modes ($\theta=\pi/2$).  This
is exactly what is expected for structures formed by gravitational
instability (\cite{Sco97,SCFFHM98}).  It would be interesting to take
advantage of this pattern of correlations to build a power spectrum
estimator that minimizes the amount of cross-correlations.

\subsection{The Power-Spectrum Covariance Matrix in Numerical Simulations}
\label{Nbody}

We measured the power spectrum from 20 different PM simulations of the
SCDM model with $\sigma_8=0.60$ and a box-size of 100~Mpc/h with
$128^3$ particles and a $256^3$ force grid.  Figure \ref{nb1}a shows
the diagonal elements of the covariance matrix normalized by the
Gaussian variance, obtained by computing the power spectra in the 20
PM simulations and performing the ensemble average.  The dashed line
at weakly non-linear scales shows the predictions of PT,
equation~(\ref{PTdiag}).  As predicted by PT, the increase in the
diagonal covariance due to non-Gaussian effects is quite small, even
at the non-linear scale $\Delta(k_{nl})=1$ (shown in the plot as a
vertical line).  We see that the diagonal terms of the covariance
matrix increase rapidly relative to the the Gaussian contributions,
once the non-linear regime is reached.

In the bottom panel we show the same diagonal components but
normalized by $P$, which yields the fractional errors in the power
spectrum estimates: $\sigma_P/P$. The fractional error decreases with
increasing $k_i$ as $k_i^{-2}$ for linear scales but does so much more
slowly once we enter the non-linear regime. It should be stressed that
the quantities shown depend both on the assumed binning in $k$-space
and on the volume of the simulation box. Unless otherwise stated, all
our measurements in numerical simulations are done using a constant bin
size, $\d k=2\pi/100$~h/Mpc, the fundamental mode of the simulation
box.  On the other hand, at sufficiently non-linear scales the
covariance matrix becomes independent of the binning (but it is still
inversely proportional to the simulation volume).

In figure \ref{nb2} we show the correlation coefficients $r_{ij}$ for
different scales.  Note that when both $k_i$ and $k_j$ fall well
within the non-linear regime, the correlation coefficient is close to
one, which implies that the power estimates at the two $k$-shells are
highly correlated.  The correlation coefficient decreases as the
shells become further apart, but it does so very slowly.  This implies
that the information-gain as more modes beyond the linear regime are
considered increases very slowly compared to the Gaussian
expectations.

Figure \ref{nb3} shows a comparison between the correlation
coefficients obtained using PT and N-Body simulations.  The comparison
is made at $\sigma_8=0.375$ ($z=0.6$), as opposed to $\sigma_8=0.60$
($z=0$) in figure~\ref{nb2}, so that more wave-modes are in the weakly
non-linear regime.  As one can see, there are large dispersions in the
measurements of $r_{ij}$ due to the limited number of N-body
simulations, but the measured values agree with the prediction of PT
within the errors.

As mentioned before, the relative importance of the Gaussian versus
non-Gaussian contributions to the covariance matrix depends upon the
choice of binning in $k$-space.  We have tested its effect with N-body
simulations and the behavior is as expected from equation~(\ref{cij}).
Note that even the distinction between diagonal and off-diagonal
covariances is somewhat arbitrary: by choosing a coarser binning of
Fourier space, what originally appears as correlations between two
nearby shells now becomes part of the diagonal variance.  As is clear
from equation~(\ref{cij}), the dependence with binning becomes less
important as we enter the non-linear regime.

Finally, figure~\ref{zspace} shows correlation coefficients in
redshift space.  These are calculated in the plane-parallel
approximation, for the same realizations shown before, in the $z=0$
output, $\sigma_8=0.60$.  The top panel shows $r_{ij}$ for the
monopole of the power spectrum, whereas the bottom panel shows
$r_{ij}$ for the quadrupole moment of the power spectrum.  Comparison
with figure~\ref{nb2} shows that the effect of correlations in
redshift space is somewhat suppressed with respect to the real-space
clustering.  That is indeed expected, since the velocity dispersion at
small scales washes out clustering and therefore non-Gaussian effects.

\subsection{The Power-Spectrum Covariance Matrix in the Non-Linear
Regime}
\label{hier}

We now study the behavior of the power-spectrum covariance matrix in
the non-linear regime, where non-Gaussian effects are most important,
and compare the numerical simulations results to those predicted by
the hierarchical model, which reproduces the observed scaling
properties of higher-order correlations in the non-linear regime.

\subsubsection{Contributions to the Covariance Matrix}
\label{4ptf}

It is instructive to write down the trispectrum in terms of the
four-point correlation function, in order to understand the nature of
the contributions to the power-spectrum covariance matrix. Writing the
connected four point function $\eta_{1234} \equiv <\d_1 \d_2 \d_3
\d_4>_c$, we have

\beq
T(\k_1, \k_2, \k_3,\k_4) = \int \frac{d^3 x_{14}}{(2\pi)^3}
\frac{d^3 x_{24}}{(2\pi)^3} \frac{d^3 x_{34}}{(2\pi)^3} \ \exp{ [-i
(\k_1 \cdot \x_{14}+\k_2 \cdot \x_{24}+\k_3 \cdot \x_{34}) ]}
\ \eta_{1234},
\eeq

\noindent where $\x_{ij} \equiv \x_i - \x_j$, and due to statistical
homogeneity $\eta_{1234}$ is only a function of three vectors, say
$\x_{14}$, $\x_{24}$ and $\x_{34}$, therefore $\k_4 =-\k_{123}$. The
trispectrum configurations relevant for the covariance matrix,
equation~(\ref{cova1}), then read

\beq
T(\k_1, -\k_1, \k_2,-\k_2) = \int \frac{d^3 x_{14}}{(2\pi)^3}
\frac{d^3 x_{24}}{(2\pi)^3} \frac{d^3 x_{34}}{(2\pi)^3} \ \exp{ [-i
(\k_1 \cdot \x_{12}+\k_2 \cdot \x_{34}) ]}
\ \eta_{1234}.
\label{T4eta}
\eeq

\noindent If we restrict to the non-linear regime, where $k_1$ and
$k_2$ are large, we see from Eq.~(\ref{T4eta}) that this restricts
$x_{12}$ and $x_{34}$ to be small; however, there is no restriction on
$x_{24}$. Therefore, the power-spectrum covariance matrix in the
non-linear regime receives contributions from all scales, not only
those in the non-linear regime. As we shall see next, this feature
takes a particular form in the context of the hierarchical model.

\subsubsection{The Hierarchical Model (HM)}

The hierarchical form for the trispectrum reads,
\beq
\label{Tha}
T(\k_1, \k_2, \k_3, \k_4) = R_a\ [P_1 P_2 P_{13} + {\rm cyc.}
] + R_b\ [P_1 P_2 P_3 + {\rm cyc.}],
\eeq

\noindent which has been proposed to explain the scaling properties of
galaxy clustering in the highly non-linear regime (\cite{FrPe78}). In
terms of the four-point function, $\eta(r)$, and the two-point
function, $\xi(r)$, the hierarchical model assumes $\eta(r) \propto
\xi^3(r)$, which implies $T \propto P^3$.  Basically, the hierarchical
model represents the trispectrum as sums of products of three power
spectra, introducing only as many parameters as there are distinct
topologies.  The $R_b$ contributions (total of $4$ terms) are the
``star'' tree-diagrams where one vertex is connected to the other
three, whereas the $R_a$ contributions are ``snake'' diagrams (total
of $12$ terms).  It is usually assumed that $R_a$ and $R_b$ saturate
to constants independent of scale and configuration in the highly
non-linear regime (and this is what we will mean by the phrase HM,
except when we discuss a modified version of HM in \S \ref{HMNbody}).
On the other hand, at large scales, PT predicts a 
hierarchical form for the trispectrum (equation~[\ref{4pt}]), but with
$R_a$ and $R_b$ strongly dependent on configuration, as illustrated in
figure~\ref{corre} and figure~\ref{Rb/Ra} below. In our case we are
interested in the configurations relevant for the power spectrum
covariance matrix, equation~(\ref{cova1}), so 4 out of the 12 $R_a$
terms vanish, because they give $P(0)=0$. This can be potentially a
problem, since in the limit that we approach these particular
configurations some scales are in the linear regime, and we do not
expect saturation (constant $R_a$ and $R_b$) to hold.  As we shall
see, this is indeed the case.

\subsubsection{The Trispectrum in the HM vs. Numerical Simulations}

Given the particular nature of the contributions to the power-spectrum
covariance matrix, we will first discuss the validity of the HM for
more general kind of configurations, were all the scales are in the
non-linear regime. We have measured the trispectrum directly in the
numerical simulations for configurations of four wave-vectors that
form a triangle, that is, $\k_1=\k_2$, and $\k_3$ at an angle $\theta$
with respect to them, for different ratios $k_3/k_1$
($\k_4=-\k_{123}$). In figure~\ref{HEPT} we present two representative
results, for $k_3/k_1=1.5$ and scales $k_1=1.26$ h/Mpc (top) and
$k_1=1.57$ h/Mpc (bottom). The results are given in terms of the
hierarchical amplitude $Q_4$ defined as

\beq
\label{Q4def}
Q_4 \equiv \frac{T_4(\k_1,\k_2,\k_3,-\k_{123})}{[P(k_1) P(k_2)
P(k_{13}) + {\rm cyc.}  ] + [P(k_1) P(k_2) P(k_3) + {\rm cyc.}]}.
\eeq

\noindent The lines show the predictions of hyperextended perturbation
theory (HEPT),

\beq
\label{hept}
Q_4^{\rm sat}(n) = (1/2)\ \frac{54-27\ 2^n+2\ 3^n + 6^n}{(1+6\ 2^n +
3\ 3^n + 6\ 6^n)} \equiv \frac{12 R_a + 4 R_b}{16}, \eeq

\noindent which gives the saturation value of the hierarchical
amplitude $Q_4$ in the highly non-linear regime in terms of the
spectral index $n=n(k)$ of the underlying linear power spectrum
(Scoccimarro \& Frieman 1998).  The spectral index has been chosen as
that of the average wave-vector at each particular configuration.
HEPT only predicts the overall amplitude of $Q_4$; in other words, it
does not attempt to model $R_a$ and $R_b$ separately. In
figure~\ref{HEPT} we show predictions for different assumptions about
the relative magnitude of $R_a$ and $R_b$, $R_a=R_b$ (solid),
$R_a=-R_b$ (dotted) and $5 R_a=R_b$ (dashed). The trend is that as
$R_b$ becomes smaller than $R_a$ (eventually becoming negative), the
configuration dependence of $Q_4$ is opposite to that observed in the
numerical simulations, where $Q_4$ is slightly enhanced at $\theta=0$
over its value at $\theta=\pi$. In the opposite case, when $R_b$
becomes larger than $R_a$, an increase of $R_b$ over $R_a$ gives a
good fit to the N-body results, up to $R_b \approx 3 R_a$. A stronger
test of the HM would require to check different type of trispectrum
configurations in addition to the ones we consider. Detailed testing
of the HM is beyond the scope of this paper, but from these results we
can conclude that at least for these type of configurations, the
trispectrum in the non-linear regime agrees very well with the HM,
with overall amplitude given by HEPT and weights $R_a \approx R_b$.

\subsubsection{The Power-Spectrum Covariance Matrix in the HM}

We now consider the calculation of the power-spectrum covariance
matrix in the HM, and show that it leads to unphysical results for the
cross-correlation coefficient in the limit that shells are widely
separated.

In order to get the bin-averaged trispectrum, equation~(\ref{cova1}),
we need to integrate over the angle between the two wave-vectors in
the two shells under consideration. It is clear that if, say, $k_2 \gg
k_1$, then $\int_{-1}^{1} (dx/2) \ P(|\k_2-\k_1|) \approx P(k_2)$. It
turns out, however, that in practice this approximation works quite
well in the non-linear regime even for $k_1=k_2$.  Thus, we get simple
approximate expressions for the diagonal and off-diagonal
contributions of the bin-averaged trispectrum

\beq
\label{HAdiag}
{\bar T}(k_i,k_i) \approx 4(2R_a+R_b)\ P_i^3.
\eeq
\beq
\label{HAndiag}
{\bar T}(k_i,k_j) \approx 2(R_a+R_b)\ P_i P_j^2 +
2(2R_a+R_b)\ P_i^2 P_j + 2R_a\ P_i^3,
\eeq

\noindent where we have assumed that $k_i> k_j$, and that $R_a$ and
$R_b$ are constants independent of configuration. These expressions
turn out to agree with results of numerical integrations over shells
to about $3\%$ for the whole range of $k_1$ and $k_2$. Note that
equation~(\ref{HAndiag}) reduces to equation~(\ref{HAdiag}) in the
limit that $k_i \rightarrow k_j$. These expressions imply that in the
limit where the non-Gaussian contribution to the covariance matrix
dominates ($k_i>k_j$)

\beq
\label{truch}
r_{ij} \approx \frac{(R_a+R_b)}{2(2R_a+R_b)} \times
\sqrt{\frac{P_j}{P_i}}.
\eeq

\noindent This result implies that, unless $R_a=-R_b$, eventually as
$k_i \gg k_j$ the cross-correlation coefficient $r_{ij}$ becomes
larger than unity, which is unphysical. As shown in figure~\ref{HEPT}
a constant relation $R_a=-R_b$ is not a good fit to the N-body results
on general trispectrum configurations.  Furthermore, it turns out that
a constant relation $R_a=-R_b$ is inconsistent with the numerical
results presented in Section~\ref{Nbody}, since it predicts that
$r_{ij}$ decays much faster with shell separation than what is
observed in the numerical simulations.  {\em We therefore conclude
that a hierarchical model with amplitudes $R_a$ and $R_b$ independent
of scale and configuration is actually not a good description of the
power-spectrum covariance matrix in the non-linear regime}.

\subsubsection{HM Predictions vs. N-body Simulations for the Covariance Matrix}\label{hmvssim}
\label{HMNbody}

Given the results in the previous sections, we expect to find
deviations from the HM predictions for the power-spectrum covariance
matrix in the non-linear regime. We now explore the nature of these
deviations by comparing with numerical simulations and use both PT and
the HM as a tool to understand the results.

The first non-trivial test is the scaling of diagonal elements. Both
PT and the HM predict that the diagonal part of the trispectrum must
scale as the third power of the power spectrum,
equations~(\ref{PTdiag})~and~(\ref{HAdiag}). Therefore, even though
the covariance matrix in the non-linear regime receives contributions
from a range of scales according to equation~(\ref{T4eta}), one
expects the HM scaling to be correct. As shown in figure~\ref{nb1}a,
the HM scaling (solid lines) agrees with the numerical simulations
(symbols). However, we find that the overall amplitude is a factor of
five times {\em smaller} than that predicted by equation~(\ref{hept})
for $R_a \approx R_b$. This suppression must come from the range of
scales (including large ones) that contribute in
equation~(\ref{T4eta}), and therefore it is
difficult to explain quantitatively without a full description of the
four-point function.

Figure~\ref{HEPT2} shows a more detailed comparison of HM predictions
with the cross-correlation coefficients $r_{ij}$ measured in the
N-body simulations.  In this calculation, we have assumed $R_a=R_b$
with $Q_4^{\rm sat}(n)$ as given by HEPT. Note the very good
agreement, especially around $k_i \approx k_j$.  As $k_i$ gets much
larger than $k_j$ we see that the predicted $r_{ij}$ start to
increase, a signature that the condition $R_a \approx R_b$ is breaking
down, as discussed above.

In light of these results, we will now use a HM with arbitrary
functions $R_a$ and $R_b$, and explore the behavior of these functions
implied by our N-body measurements. Note that once we give up the
hypothesis that $R_a$ and $R_b$ are strictly constants, we should
distinguish between two kinds of $R_a$ and $R_b$: the $R_a$ and $R_b$
in equation (\ref{Tha}) which depend on both configuration and scale,
and the effective $R_a$ and $R_b$ in equations
(\ref{HAdiag}-\ref{HAndiag}) which have the configuration dependence
integrated out.  We will rely on the context to differentiate between
the two meanings here.

Given our measurements of the power spectrum covariance matrix
presented in Section~\ref{Nbody}, we can obtain estimates for $R_a$
and $R_b$ from equations (\ref{HAdiag}-\ref{HAndiag}). However, since
the largest contribution is due to the first term in
equation~(\ref{HAndiag}), the best constraint is on the average value,
$(R_a+R_b)/2$.  This is shown in figure~\ref{RaRb}, for the particular
case of $k_j=0.98$~h/Mpc as a function of $k_i$. We see that the
average value shows a systematic decrease with shell separation.
Since this configuration dependence is likely to come from scales not
in the non-linear regime, it is useful to consider PT, which predicts
a specific relation between $R_a$ and $R_b$ as a function of shell
separation, to see whether we can explain the decline in
figure~\ref{RaRb}.

In figure~\ref{Rb/Ra} we show the ratio $R_b/R_a$ predicted by PT as a
function of configuration angle $\theta$ between $\k_1$ and $\k_2$
for different scales.  We see that for shells close to each other,
although the ratio $R_b/R_a$ varies, on average $R_a \approx R_b$ (in
fact, for $k_1=k_2=1$~h/Mpc, the angular average yields $R_a \approx
1.5 R_b$).  However, as the shells become more separated, the averaged
relation between $R_a$ and $R_b$ changes, reaching $R_a=-R_b$ in the
limit of large separation, just what is needed in a hierarchical model
to preserve the condition that the cross-correlation coefficient
$r_{ij} \leq 1$.

We now examine whether the type of configuration dependence valid
in PT can explain the behavior of $(R_a+R_b)/2$ deduced from the
N-body results. In figure~\ref{RaRb}, the solid line shows the
prediction for $(R_a+R_b)/2$ as a function of shell separation
assuming that the {\em averaged} relation between $R_a$ and $R_b$
is the same as that given by PT. The overall amplitude has been
chosen to fit the diagonal values, at $k_i=k_j=0.98$~h/Mpc.  We
see that this prediction works quite well to explain the
configuration dependence of the covariance matrix in the
non-linear regime.

\subsubsection{Summary}

To summarize, we found that the hierarchical model (HM) is not a
valid description of the power-spectrum covariance matrix in the
non-linear regime. Instead, we find that the covariance matrix
behaves as a ``generalized'' HM with a configuration dependence
that resembles that in PT. We interpret this as a consequence of
the fact that even in the non-linear regime the covariance matrix
receives contributions from large scales.  On the other hand, we
have tested the HM for other trispectrum configurations in the
non-linear regime and found good agreement for $R_a \approx R_b$
and overall amplitude given by HEPT.

Although N-body results have been restricted to the SCDM model, it is
worth mentioning that hierarhical amplitudes such as $Q_4$, $R_a$ and
$R_b$ are expected to be very insensitive to cosmological parameters,
as can be shown quite generally from the equations of motion of
gravitational instability (Scoccimarro et al. 1998, Appendix
B.3). Therefore, we expect that the dependence of the covariance
matrix on cosmology will appear only through the overall dependence on
the power spectrum.

\section{Impact on Cosmological Parameter Determination}
\label{parameter}

To assess the importance of the non-Gaussian terms in the covariance
matrix, suppose we are interested in only one parameter: the amplitude
of the power spectrum over a range of scales.  Let us assume that the
shape, $\bar P(k_i)$, is known and let us estimate the amplitude
$x=P(k_i)/\bar P(k_i)$.  If only the Gaussian terms of the covariance
were important, the minimum variance estimator for the amplitude would
be,
\begin{eqnarray}
\hat x= {\sum_{k_i < k_{\rm max}} N_{k_i} \hat P(k_i)/\bar P(k_i)\over
\sum_{k_i
< k_{\rm max}} N_{k_i} },
\end{eqnarray}
where we simply weigh the power in each bin by its inverse variance,
and all bins are used up to some $k_{\rm max}$. In the same Gaussian
limit the variance of $\hat x$ would be
\begin{eqnarray}
\sigma_x^2=\langle \hat x^2 \rangle-\langle \hat x\rangle^2 ={2\over
N_t},
\end{eqnarray}
where $N_t=\sum N_{k_i}$ is the total number of $\bi k$ modes used.

Figure \ref{PT3} shows the ratio $\sigma_x^2/(2/N_t)$ obtained using
perturbation theory for a 1~Gpc/h box.  On large scales, $k_{\rm max}
\, \approxlt \, 0.1$~h/Mpc, PT shows that the non-Gaussian
contributions do not have a significant impact on the determination of
$x$, the error-bar starts increasing rapidly thereafter (for this
cosmological model $k_{nl}=0.33$~h/Mpc, and at the smoothing scale of
$R \sim 1/k_{nl}$, $S_4 \sigma^2 / 5$ is of order 1).  In
figure~\ref{nb4}a, we show the corresponding results from N-body
simulations.  In figure~\ref{nb4}b we show the effective number of
modes, defined as $N_{\rm eff}\equiv 2/\sigma_x^2$.  As we can see,
$N_{\rm eff}$ varies much more slowly once we enter the non-linear
regime.

In the non-linear regime, different band-power estimates become highly
correlated.  What this means in practice is that the power at
different bands fluctuate up and down together, making the overall
amplitude of the power spectrum difficult to determine while
essentially preserving information on its shape. This can be made more
precise by considering the following covariance matrix:
\begin{equation}
C_{ij} = \beta [P_i P_j]^{3\over 2} (1-\epsilon_{ij})
\label{cijnonlinear}
\end{equation}
which follows from equation (\ref{cij}), ignoring the Gaussian term,
and from equation (\ref{HAdiag}).  The parameter $\beta$ is some
simple constant proportional to $V_f$.  The symmetric matrix
$\epsilon_{ij}$ has vanishing diagonal elements, $\epsilon_{ii} = 0$,
and the off-diagonal elements determine the cross-correlations between
different band-powers.  We will consider the limit in which
$\epsilon_{ij}$ is small (see figure \ref{HEPT2}), in other words all
modes are highly correlated and $C_{ij}$ is nearly singular.
Furthermore, as is clear from figure \ref{HEPT2}, $\epsilon_{ij}$ is
slowly varying, and we will treat $\epsilon_{ij} = \epsilon$ as a
constant.

Let us consider the following rescaling of the power spectrum
estimate:
\begin{eqnarray}
\hat y_i={\hat P_i \over {{\sqrt\beta} [P_i]^{3/2}}}.
\label{hatydef}
\end{eqnarray}
where the true power spectrum $P(k_i)$ is of course unknown a priori,
but one can think of it instead as a fiducial power spectrum and we
are trying to measure deviations from it.  (It is actually not
necessary to make this change of variable at all, but it will simplify
some of our expressions below.)  The corresponding covariance matrix
is then
\begin{equation}
C^y_{ij} = \langle \hat y_i \hat y_j \rangle - \langle \hat y_i
\rangle \langle \hat y_j \rangle = 1 - \epsilon\ (1-\delta_{ij})
\label{Cyij}
\end{equation}
We can diagonalize this matrix and ask which combinations of the band
powers are better determined (see \cite{Hamilton97b} for other choices
of rescaling and diagonalization).  There are two types of eigenvalues.
The first is $\lambda_1=N - (N-1) \epsilon$, associated with the
eigenvector ${\bi e}_1=(1,1,1,....,1)/\sqrt{N}$, where $N$ is the
number of bands.  This eigenvector corresponds to the overall
amplitude of the rescaled power spectrum.  In addition, there are
$N-1$ degenerate eigenvectors, each with an eigenvalue $\lambda
=\epsilon$.  The corresponding eigenvectors are ${\bi
e}_m=(0,0,...1,-1,0...,0)/\sqrt{2}$, where the $1$ is in the $m-1$-th
position ($m=2,...,N$).  Each of these eigenvectors provides a measure
of the local derivative of the rescaled power spectrum.  Note that the
above set of eigenvectors are linearly independent but do not form an
orthogonal basis.  In the case where $\epsilon$ (or more precisely
$\epsilon_{ij}$) actually varies slowly with the band, but still
remains small, the above conclusions remain largely unchanged:
$\lambda_1 = N + O(\epsilon_{ij})$, and the rest of the eigenvalues
satisfy $\lambda \sim \epsilon$ ($\epsilon_{ij} \sim \epsilon$ plus
small variations) except that the degeneracy is lifted by the slowly
varying $\epsilon_{ij}$.

Clearly, $\lambda_1$ is much larger than any other eigenvalues.
Hence, the combination corresponding to ${\bi e}_1$ is the most poorly
determined.  To be more precise, the fractional error in the quantity
$\bi {\hat y} \cdot \bi e_1$ is
\begin{equation}
\sqrt{\langle [\bi {\hat y} \cdot \bi e_1 - \langle \bi {\hat y} \cdot \bi
e_1 \rangle]^2 \rangle \over  \langle \bi {\hat y} \cdot \bi e_1
\rangle^2} = (1 + O(\epsilon)) \left[ {1\over N} \sum_i {1\over
\sqrt{\beta P_i}} \right]^{-1} \, .
\label{errorE1}
\end{equation}
To the extent that $\sqrt{P(k_i)}$ is slowly varying with $k_i$, the
fractional error is approximately independent of $N$, which means
increasing the number of bands does not help reduce the error in this
estimate of the amplitude of the rescaled power spectrum.  The only
way to reduce the error is to make $\beta$ smaller, i.e.  having a
larger survey volume.

On the other hand, the local spectral index of the rescaled power
spectrum can be defined as $\bi {\hat y} \cdot \bi e_m / \langle \hat
y_m \rangle$ ($m \ne 1$), and its error is given by (we use error
instead of fractional error here because the local spectral index
could vanish):
\begin{equation}
{\sqrt{\langle [\bi {\hat y} \cdot \bi e_m - \langle \bi {\hat y} \cdot \bi
e_m \rangle]^2 \rangle} \over  \langle \hat y_m \rangle} =
\sqrt{\epsilon \beta P_m}
\end{equation}
which is $O(\sqrt{\epsilon})$ smaller than the error in the rescaled
amplitude.  It is intriguing that even though only two neighboring
bands are required to estimate the shape while all bands are used to
estimate the amplitude, it is the former that is better determined.
This is a result of the high correlation limit that we have taken,
which makes averaging over band-powers nearly useless.

\section{Covariance Matrix of the Projected Power Spectrum:
Application to Weak Lensing}
\label{lensing}

We shall now consider the statistics of the (2D) projected density
field.  Suppose one observes a small patch of the sky covering a solid
angle $\Omega$.  The patch is taken to be small enough that we can use
the flat sky approximation and expand the projected mass density in
Fourier modes instead of spherical harmonics.  We will consider
\begin{equation}
\kappa(\bi \theta)=
\int_0^{\chi_0} d\chi\ w(\chi)\ \delta(r(\chi)\theta_x,r(\chi)\theta_y,\chi),
\label{proydef}
\end{equation}
where $w(\chi)$ is a slowly varying function of the radial comoving
distance to the observer $\chi$.  We introduce the angular-diameter
distance $r(\chi)$, $r(\chi)=K^{-1/2}\sin K^{-1/2}\chi, \chi,
(-K)^{-1/2}\sinh (-K)^{-1/2}\chi$ for models with positive, zero and
negative curvature respectively, and $K=(\Omega_0-1)H_0^2$ with $H_0$
the present value of the Hubble constant.  For weak-lensing the
relevant weight function $w(\chi)$ is
\begin{eqnarray}
w(\chi)&=&{3\over 2} H_0^2 \Omega_m  {g(\chi_0,\chi) \over a}
\nonumber \\
g(\chi,\chi^\prime)&=&{r(\chi^\prime)\ r(\chi-\chi^\prime)
\over r(\chi)}\, ,
\end{eqnarray}
in which case $\kappa$ is the weak-lensing convergence for background
galaxies located at $\chi = \chi_0$. Note that our definition of
$\kappa$ here is a factor of 2 larger than the one commonly used (see
e.g. Jain et al. 1999). We are interested in the Fourier components of
the projected field, from equation~(\ref{proydef}) we obtain

\begin{equation}
\kappa(\bi l)= \int {d^2 {\bi \theta}\over (2\pi)^2}\ e^{-i{\bi
 l}\cdot {\bi \theta}} \ \kappa(\bi \theta) = \int d\chi
 \frac{w(\chi)}{r^2(\chi)} \ \int_{-\infty}^\infty dk \ \d\Big(\frac{\bi
 l}{r(\chi)},k\Big)\ e^{i k \chi},
\label{kappa_l}
\end{equation}
which gives the two-point correlator

\beq
\lexp \kappa(\bi l) \kappa(\bi l') \rexp = \int d\chi_1 \
\frac{w_1}{r_1^2} \int d\chi_2 \ \frac{w_2}{r_2^2}\ \dD\Big( \frac{\bi
l}{r_1}+\frac{\bi l'}{r_2}\Big) \times \int_{-\infty}^\infty dk\
e^{i k (\chi_1-\chi_2)}\ P[\sqrt{(l^2/r_1^2)+k^2}]\,
\label{kappa2}
\eeq

\noindent where $w_i \equiv w(\chi_i)$ and similarly $r_i \equiv
r(\chi_i)$. The integral over the line-of-sight wave-vector is
dominated by $k (\chi_1-\chi_2) \equiv k \Delta\chi \la 1$, where
$\Delta\chi$ must be the typical scale of variation of
$w(\chi)/r(\chi)^2$ (otherwise the total integration vanishes). In the
small angle approximation, it is reasonable to assume that over the
scales of interest $w(\chi)/r(\chi)^2$ is approximately constant, that
is, $l/r \gg 1/\Delta\chi$. Therefore, $l^2/r_1^2 \gg k^2$, i.e. only
perpendicular Fourier modes contribute to the projected field, then
$P(\sqrt{l^2/r_1^2+k^2}) \approx P(l/r_1)$ and the integral over $k$
gives a delta function in $\Delta\chi$. Thus, we obtain

\beq
\lexp \kappa(\bi l) \kappa(\bi l') \rexp = (2\pi)\ \dD({\bi l}+{\bi
l'}) \
\int d\chi \ \frac{w^2(\chi)}{r^2(\chi)}  \ P[l/r(\chi)],
\label{kappa3}
\eeq
which is nothing but Limber's equation
(\cite{Peebles80,Kaiser92,Kaiser98}).  The power spectrum estimator
for the projected density field is accordingly

\begin{eqnarray}
{\hat P}_\kappa(l_i)&=& A_f \int_{l_i} {d^2{\bi l}\over
A_r(l_i)}\ \kappa({\bi l})\kappa(-{\bi l}),
\label{estimpk}
\end{eqnarray}
where $A_f=(2\pi)^2/\Omega$ is the area of the fundamental cell in
$\bi l$ space, and we have assumed an average over a ring in $\bi
l$-space with area $A_r(l_i)$ and centered at $l_i=i\times \delta
l$. Equations (\ref{kappa3}) and (\ref{estimpk}) lead to

\begin{eqnarray}
\langle \hat P_\kappa(l) \rangle
&=&(2\pi) \int d\chi {w^2(\chi)\over r^2(\chi)} P[l/r(\chi)].
\label{limber}
\end{eqnarray}
In figure \ref{lensing1}a we
show the power spectrum for the weak-lensing convergence assuming that
the background galaxies are at a redshift of $z=1$.  Following Jain \&
Seljak (1997), we use a model of the non-linear power spectrum
proposed by Hamilton et al. (1991), and later extended by Peacock \&
Dodds (1994,1996) and Jain et al. (1995).  In panel~b we show the
integrand of equation (\ref{limber}) as a function of redshift for
$l=1000$ and $l=10000$.  The area under the curve is proportional to
$P_\kappa$.  We conclude that the peak contribution comes from
$z=0.3-0.4$ depending on the $l$ of interest, but that the integrand
is a very broad function of redshift.

The previous calculation can be easily extended to higher moments of
the projected field.  We obtain

\begin{eqnarray}
\langle \kappa(\bi l_1)\cdots \kappa(\bi l_n)\rangle_c
= (2\pi)^{n-1} \dD(\bi l_1+\cdots +\bi l_n)\
\int d\chi\ {w^n(\chi)\over
r^{2n-2}(\chi)}\ T_n[\bi l_1/r(\chi),\cdots,
\bi l_n/r(\chi)].
\label{higherpr}
\end{eqnarray}
Equations (\ref{estimpk}) and (\ref{higherpr}) can be used to
calculate the covariance matrix of the power spectrum estimator,
\begin{eqnarray}
C_{ij}&=& A_f\ [{2P_\kappa^2(l_i) \over A_r(l_i)}\delta_{ij}+
\bar T_\kappa(l_i, l_j)] \nonumber \\
& &  \nonumber \\
\bar T_\kappa(l_i, l_j) &\equiv&
(2\pi)^3 \int d\chi {w^4(\chi) \over r^6(\chi)} \int_{l_i} {d^2 {\bi l}_1
\over A_r(l_i)} \int_{l_j} {d^2 {\bi l}_2 \over A_r(l_j)} T[\bi
l_1/r(\chi),-\bi l_1/r(\chi),\bi l_2/r(\chi),-\bi l_2/r(\chi)].
\label{covproy}
\end{eqnarray}

Before presenting the results of the numerical evaluation of the
different terms in equation (\ref{covproy}), we will make an order of
magnitude estimate of the size of the different terms.  In the
process, we hope to gain some physical insight and understand how the
different terms scale with the parameters of the problem.  The ratio
of the non-Gaussian to Gaussian terms is (for $i = j$)
\begin{eqnarray}
{\cal R}={A_r \bar T_\kappa \over 2 P_\kappa^2}.
\end{eqnarray}
We then make the following crude approximations:
\begin{eqnarray}
P_\kappa(l)&\approx& (2\pi) \Delta\chi^*\ {w^2(\chi^*) \over r(\chi^*)^2}
\ P[l/r(\chi^*)]\, ,
\nonumber \\
\bar T_\kappa(l,l)&\approx& (2\pi)^3 \Delta\chi^*\
{w^4(\chi^*) \over r(\chi^*)^6}\
\bar T[l/r(\chi^*),l/r(\chi^*)]\, ,
\end{eqnarray}
where we have assumed that the line of sight integral is dominated by
contributions from $\chi^* \pm \Delta \chi^*/2$, and that the
integrand remains roughly constant over this interval.  We neglect the
configuration dependence of the trispectrum and just use its typical
value at the scale of interest.  Note that both the power spectrum and
the trispectrum are a function of time so they should be evaluated at
$\chi^*$.  Under these simplifying assumptions the ratio becomes
\begin{eqnarray}
{\cal R}={\pi A_r \over r^2(\chi^*) \Delta\chi^*} {\bar T \over
P^2}.
\label{Rww}
\end{eqnarray}

It is interesting to compare these formulae with their analogues in 3D
(equation [\ref{cij}]).  If we make the identification $k =
l/r(\chi^*)$, $\Delta k_\chi = 2\pi / \Delta \chi^*$ and $\delta k =
\delta l /r(\chi^*)$, we can rewrite the prefactor ${\pi A_r /
r^2(\chi^*) \Delta\chi^*}$ as $2 \pi k \Delta k_\chi \delta k / 2$.
Let's compare this with the relevant prefactor in the 3D case: $4 \pi
k^2 \delta k / 2$. By assumption, $k \gg \Delta k_\chi$, which means
that the Gaussian variance is much larger, or the ratio of
non-Gaussian to Gaussian terms is much smaller, in 2D compared to 3D.
This is a result of the fact that only the Fourier modes perpendicular
to the line of sight contribute to the 2D projection, by virtue of an
otherwise rapidly oscillating integrand (see derivation above).  This
means a far fewer number of non-linear modes are available in 2D
compared to 3D, for a given $k$.  While in 3D all the modes in a shell
of volume $4\pi k^2\delta k$ are available, in 2D only those in a
ring-shaped region of area $2\pi k\delta k$ and height $\Delta k_\chi$
contribute to non-Gaussianity.  Hence, projection raises the
significance of the Gaussian variance relative to the non-Gaussian
term.

The projection has another interesting effect.  As discussed in
the context of the hierarchical ansatz, the trispectrum scales as
the third power of the power spectrum.  Contrary to what happens
for the three-dimensional case, the number of modes only increases
like $k^2=l^2/r^2(\chi)$ as we go to higher $l$ (we are assuming
$\delta l \sim l$ here as it gives the correct order of magnitude
estimate of the true significance of non-Gaussian terms). The
ratio $\cal R$ scales as ${\cal R}\propto {k^2 P(k)/\Delta\chi^*}$
instead of $k^3 P(k)$ for the 3D case.  As a consequence, if the
3D power spectrum decreases faster than $k^{-2}$, the relative
importance of the non-Gaussian variance can actually decrease as
one goes to smaller angular scales.

As an example we consider the weak-lensing effect on background
galaxies at a redshift of $z\sim 1$. We then take $\chi^*$
corresponding to a redshift of $z\sim 0.4$, which means a distance of
$r(\chi^*)=\chi^* \sim 1 {\, \rm h^{-1} Gpc}$.  Because the integrand
is a slowly varying function of $\chi$ we will take $\Delta \chi^*
\sim \chi^*$.  We will concentrate on scales $l>1000$ ($\theta< 1/ l
\approx 3^{'}$ in the correlation function) which at this distance
correspond to $k>1$~h/Mpc.  Since these scales are all in the
non-linear regime, to estimate the magnitude of $\cal R$ we use the
value of the trispectrum obtained in the simulations, ${\bar T}\sim 17
P^3$, to get
\begin{eqnarray}
{\cal R}\sim 0.2\ \Big({l \over 1000}\Big)\ \Big({\delta l \over
1000}\Big) \Big({1 {\, \rm h^{-1} Gpc} \over r(\chi)}\Big)^2 \Big({1
{\, \rm h^{-1} Gpc} \over \Delta\chi}\Big) \Big({\bar T/P^3 \over 17}\
{P(l/1 {\, \rm h^{-1}Gpc}) \over {0.5 {\, \rm h^{-3} Mpc^3}}}\Big)
\label{simpleest2}
\end{eqnarray}
This is in contrast with the 3D case where the corresponding $\cal R$
is larger than $1$ for $k>1$~h/Mpc.  As we said above, projection
decreases the effective number of non-linear modes which makes the
Gaussian variance comparatively more important.

We can compare our estimate of $\cal R$ with other measures of
non-Gaussianity in simulations of weak lensing. Jain et al. (1999)
computed the skewness of $\kappa$ and found $S_3\sigma\approx 1.6$ on
a scale of $3^{'}$. To make the connection with our result we will
assume that this number also estimates $S_4\sigma^2$, as expected by
simple scaling. Then we note that $S_4\sigma^2$ can be written as,
\begin{equation}
S_4(R)\sigma^2=\frac{\int d^2l_1 d^2l_2 d^2l_3 \ W_1 W_2 W_3
W_{123}\ T_\kappa(\bi l_1,\bi l_2,\bi l_3,-\bi l_{123})}{
 [\int d^2l \ W^2(lR)\ P_\kappa(l)]^2},
\label{simpleest}
\end{equation}
where $W_i \equiv W(l_i R)$, with $W$ being the Fourier transform of
the smoothing window. We see that $S_4\sigma^2$ measures the same
ratio of trispectrum and power spectrum that is relevant for $\cal R$
but with one caveat. $S_4\sigma^2$ is determined by an average over
all trispectrum configurations while $\cal R$ is sensitive to those
particular shapes relevant for the covariance matrix, and as we
discussed in \S \ref{hmvssim}, the latter is smaller than the naive
expectation by a factor of five. Thus we expect $S_4\sigma^2/5$ to be
comparable to $\cal R$, as indeed it is.

Also note that the weight function $w(\chi)$ cancels out in the ratio
$\cal R$, so the above conclusions apply equally well to other types
of projection, such as that in angular galaxy surveys.  The
non-Gaussian contribution to the error is comparable but does not
dominate over the Gaussian contribution for the cluster-normalized
SCDM model, and for mass distributions projected over cosmological
distances ($\sim 1 {\, \rm h^{-1} Gpc}$). For angular surveys,
however, the effects of non-Gaussianity are likely to be larger than
in the weak-lensing case, since the projection takes place over a
smaller range of scales.

In figure \ref{lensing2} we show the result of numerically integrating
equation (\ref{covproy}).  We use here the hierarchical form together
with HEPT discussed in \S \ref{hier}, but rescale the amplitude of
$R_a$ ($\approx R_b$) to match the amplitude of the power spectrum
variance measured in the simulations ($T/P^3=17$ at $k = 1 {\, \rm {h
Mpc^{-1}}}$ for the SCDM model). The figure shows the ratio of the
diagonal terms of the covariance so this value is all that is needed
to normalize the calculation.  This should be a good approximation for
most of the scales shown, but would start to break down close to $l
\sim 100$, in which case the PT result would begin to take over (see
equation \ref{PTdiag}).  We show the results for three cosmological
models, the SCDM model we have been considering so far
($\sigma_8=0.60,\ \Gamma=0.5$), an open model with $\Omega_m=0.3$,
$\Gamma=0.21$ and $\sigma_8=0.85$, and a flat cosmological constant
($\Lambda$) dominated model with $\Omega_m=0.3$ and
$\Omega_\Lambda=0.7$.  We have assumed the the hierarchical ratios
$R_a$ and $R_b$ are independent of the cosmological model, which
should hold to a very good approximation (\cite{SCFFHM98}).

The ratio $\cal R$ obtained in the numerical integration for SCDM is
in agreement with our simple estimate in equation (\ref{simpleest2}).
Furthermore we can see that the ratio has a maximum around $l\approx
8000$, which is a consequence of the fact that the 3D power spectrum
decreases faster than $k^{-2}$ therefater.  Even though we are probing
scales that are more non-linear as we go to higher $l$, the projection
actually makes the estimates of the power spectrum of $\kappa$ more
Gaussian, since the number of non-linear modes does not increase as
fast as to compensate for the decaying power spectrum. Therefore,
contrary to what happens in 3D, the non-Gaussian effects never
dominate even at very small scales for the SCDM model.

Figure \ref{lensing2} shows that the non-Gaussian effects can be more
important in other models, however.  The main difference between
models arises due to their different normalizations, parametrized by
$\sigma_8$ (in all cases, the normalization is chosen to yield the
correct cluster abundance today).  An additional effect which enhances
the signature in the open model is the difference in the fluctuation
growth rates.  The parameter $\sigma_8$ specifies the normalization at
the present time but the lensing effect is sensitive to the amplitude
of the fluctuations all the way up to the redshift of the background
galaxies.  In the open model structure grows slower than in the flat
one, so for a given normalization today the average power up to a
redshift $z\sim 1$ is significantly larger.  The growth rate of the
cosmological constant model is intermediate between that of the open
and flat models. Hence, although the open and cosmological constant
models have the same present normalization, the size of the ratio in
figure~\ref{lensing2} differs for the two models. Finally, there is
one more differentiating factor: the angular-diameter distance to the
relevant redshifts is largest in the $\Lambda$ model, which tends to
further reduce $\cal R$ (equation [\ref{simpleest2}]).

Our results for different cosmologies agree roughly with the numerical
investigations carried out by van Waerbeke et al.  (1998). We
generally find, however, a smaller (by a factor of order two)
non-Gaussian contribution than they do. The source of this discrepancy
is likely the different treatment of the dynamics of gravitational
clustering. As described above, we use as an approximation the
hierarchical model for the diagonal non-Gaussian contribution; on the
other hand, van Waerbeke et al. (1998) use the dynamics of
second-order Lagrangian perturbation theory in 2D.

In summary, we have shown that the non-Gaussian terms in the diagonal
of the projected power spectrum, which arise due to the non-linear
nature of gravitational clustering, can be the dominant source of
error in some models, depending on the normalization and cosmology.
These terms should be included when analyzing future surveys or when
trying to predict their capabilities (\cite{VWBM98,HuTe98}). Perhaps
more importantly, as in the three-dimensional case, gravitational
clustering induces correlations between different band-powers, in
addition to increasing their individual error-bars.  Calculations
similar to those performed for figure \ref{HEPT2} are straightforward
to carry out for weak-lensing, but will not be included here.

\section{Discussion}
\label{discuss}

\begin{figure*}
\begin{center}
\leavevmode
\epsfxsize=4.0in \epsfbox{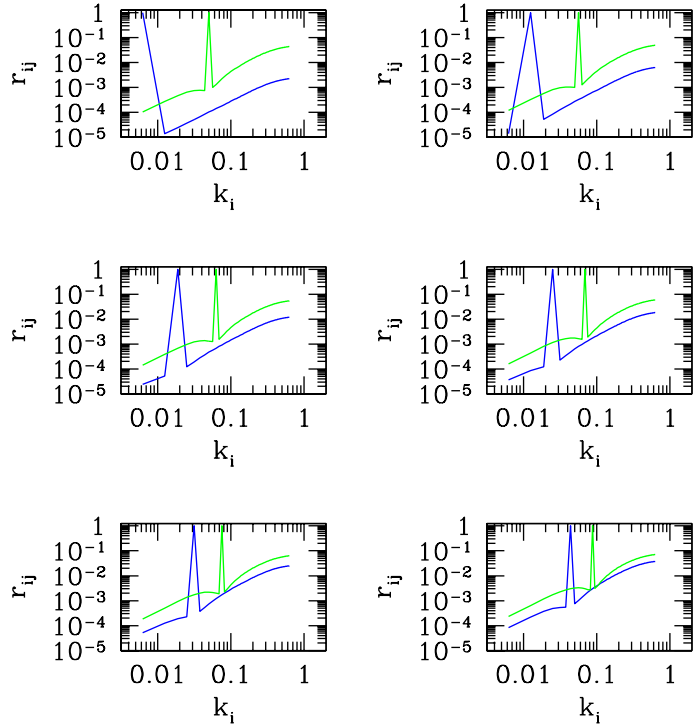}
\end{center}
\caption{Correlation coefficients $r_{ij} \equiv C_{ij} / [C_{ii}
C_{jj}]^{1/2}$ as a function of $k_i$ for different fixed $k_j$
obtained using PT for a 1 Gpc/h box, and the SCDM model.  The $k$
shells are the same as in figure \ref{PT1}.  The values of $k_j$ can
be inferred from the place where $r_{ij}=1$.  In each panel we plot
the results for two different values of $k_j$.}
\label{PT2}
\end{figure*}

\begin{figure*}
\begin{center}
\leavevmode
\epsfxsize=4.0in \epsfbox{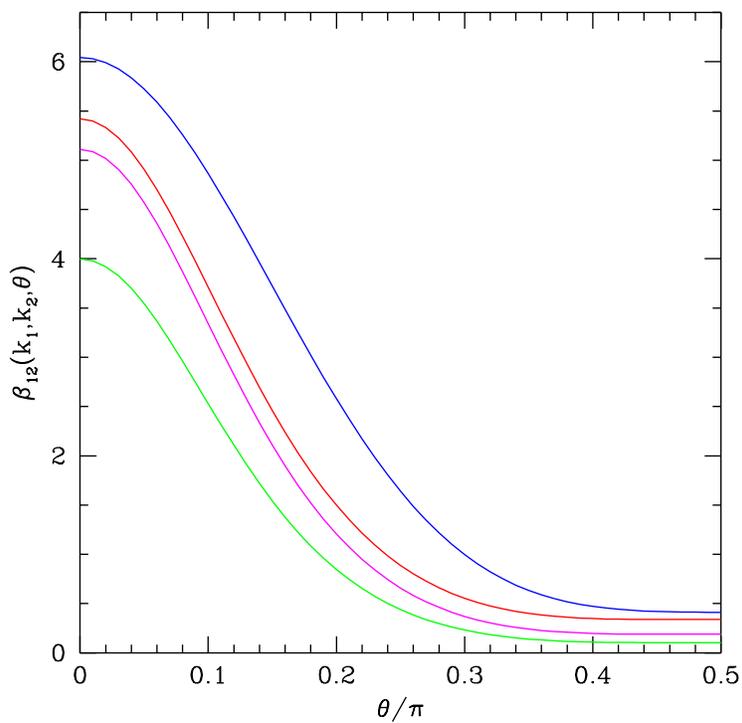}
\end{center}
\caption{The coefficient $\beta_{12}(k_1,k_2,\theta)$ [equation~(\protect
\ref{beta})] as a function of $\theta$ for different scales. From top
to bottom, $(k_1,k_2)=(0.1,0.4)$~h/Mpc, $(0.1,0.2)$~h/Mpc,
$(0.2,0.4)$~h/Mpc, and $(0.4,0.8)$~h/Mpc.}
\label{corre}
\end{figure*}

\begin{figure*}
\begin{center}
\leavevmode
\epsfxsize=4.0in \epsfbox{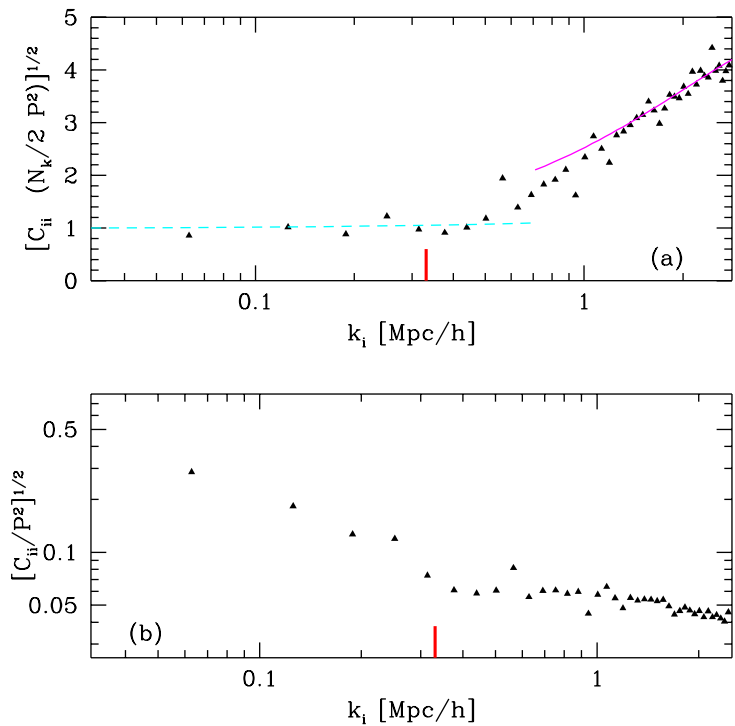}
\end{center}
\caption{The top panel shows the diagonal elements of the covariance
matrix normalized by the Gaussian variance, obtained by comparing 20
PM simulations.  The dashed line shows the predictions of PT,
equation~(\protect\ref{PTdiag}), and the solid line the hierarchical
scaling, equation~(\protect\ref{HAdiag}).  The bottom panel shows the
fractional error in the band-power estimates.  This fractional error
scales with the size of the survey or simulation box, the results in
the figure correspond to a volume $V_0=(100 {\, \rm h^{-1}Mpc})^3$.
Results for other volumes can be obtained by scaling by
$(V_0/V)^{1/2}$.  The vertical line on the $x$ axis indicates the
non-linear scale.  The width of shells in $k$-space is $\d k=2\pi/100$
h/Mpc.}
\label{nb1}
\end{figure*}

\begin{figure*}
\begin{center}
\leavevmode
\epsfxsize=4.0in \epsfbox{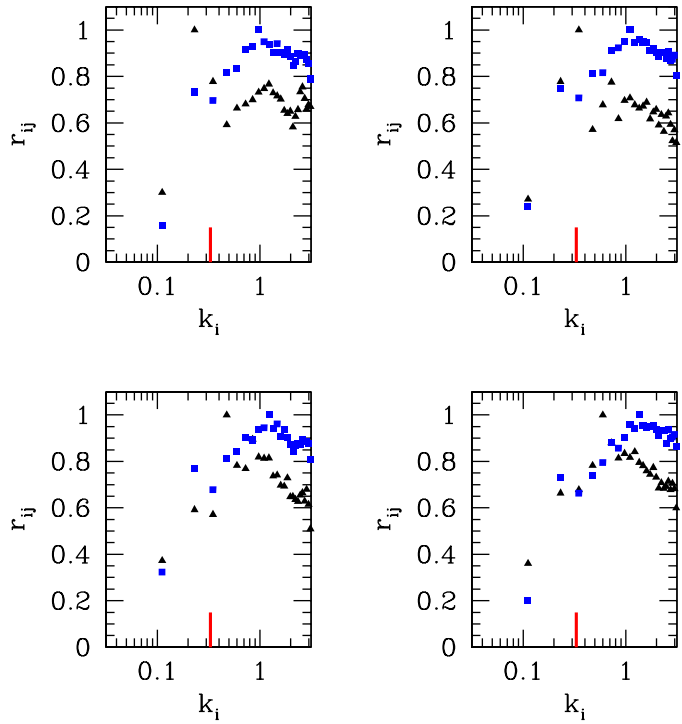}
\end{center}
\caption{Correlation coefficients $r_{ij}$ from the N-body simulations
as a function of $k_i$ for a fixed $k_j$.  The two sets of points
(triangles and squares) in each plot correspond to different $k_j$'s.
The different values of $k_j$ can be deduced from the place where
$r_{ij}=1$.  The mark on the $x$ axis indicates the non-linear scale.
The width of shells in $k$-space is $\d k=4\pi/100$ h/Mpc.}
\label{nb2}
\end{figure*}

\begin{figure*}
\begin{center}
\leavevmode
\epsfxsize=4.0in \epsfbox{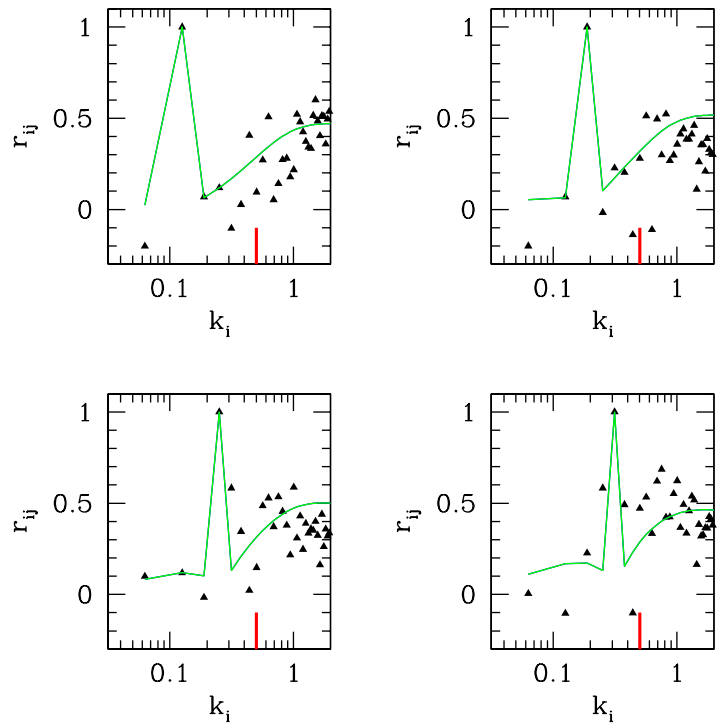}
\end{center}
\caption{A comparison of $r_{ij}$ obtained from PT and N-body
simulations, at $\sigma_8=0.375$ ($z=0.6$).  Note the noise level in
the covariance matrix measured from the numerical simulations; this is
due to the low number (20) of realizations used to estimate the
ensemble averages.  The width of shells in $k$-space is $\d
k=2\pi/100$ h/Mpc.}
\label{nb3}
\end{figure*}

\begin{figure*}
\begin{center}
\leavevmode
\epsfxsize=4.0in \epsfbox{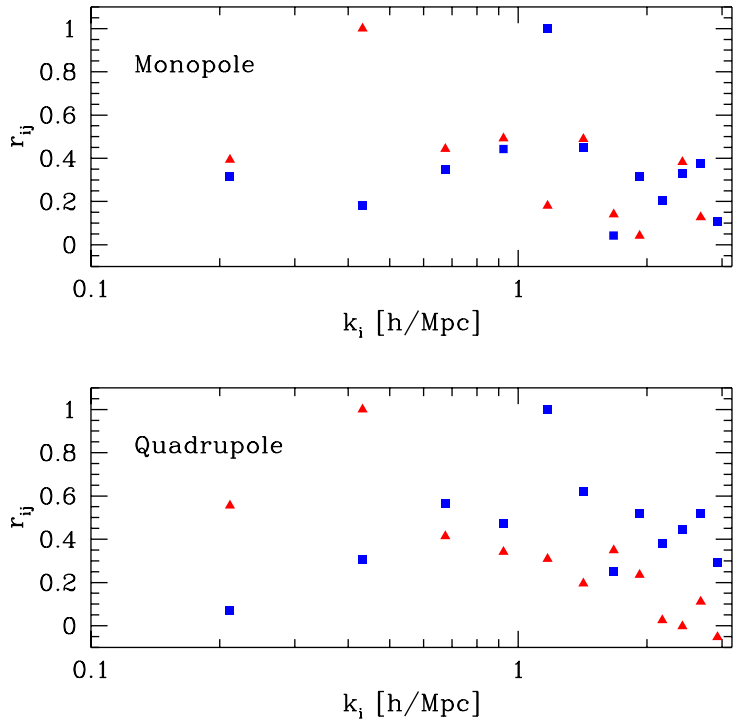}
\end{center}
\caption{Cross-correlation coefficients $r_{ij}$ obtained N-body
simulations in redshift space, for $\sigma_8=0.60$.  The top panel
shows $r_{ij}$ for the monopole of the power spectrum, whereas the
bottom panel shows $r_{ij}$ for the quadrupole power spectrum.  The
width of shells in $k$-space is $\d k=8\pi/100$ h/Mpc.}
\label{zspace}
\end{figure*}

\begin{figure*}
\begin{center}
\leavevmode
\epsfxsize=4.0in \epsfbox{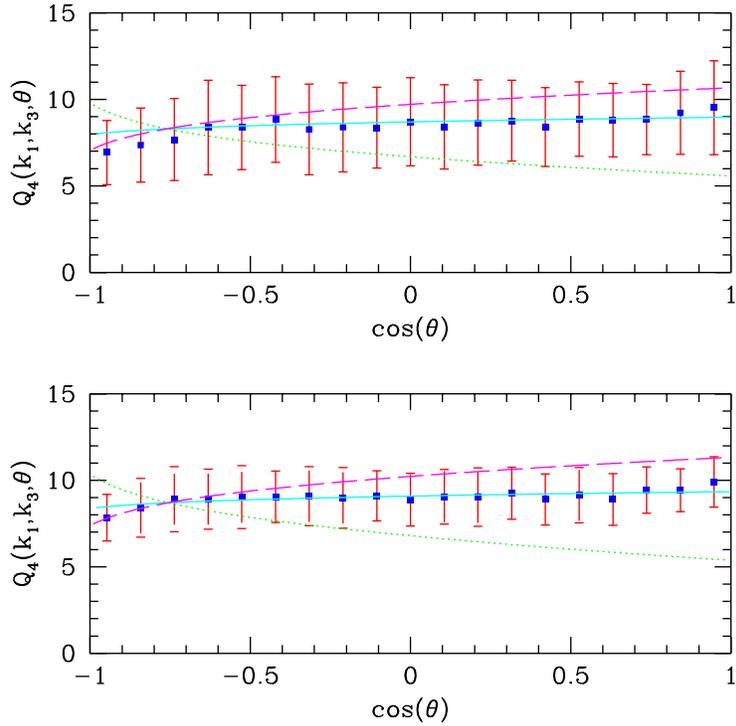}
\end{center}
\caption{The trispectrum hierarchical amplitude $Q_4$ for
configurations $\k_1=\k_2$ and $k_3 =1.5 k_1$ as a function of the
angle $\cos (\theta) \equiv {\hat k_1} \cdot {\hat k_3}$ in the
non-linear regime, for two different scales, $k_1=1.26$ h/Mpc
(top) and $k_1=1.57$ h/Mpc (bottom).  The lines show the
predictions of HEPT (equation \protect\ref{hept}) plus assuming
$R_a=R_b$ (solid), $R_a=-R_b$ (dotted) and $5 R_a=R_b$ (dashed).
The numerical simulations results are denoted by symbols with
error bars determined from 20 realizations.  } \label{HEPT}
\end{figure*}

\begin{figure*}
\begin{center}
\leavevmode
\epsfxsize=4.0in \epsfbox{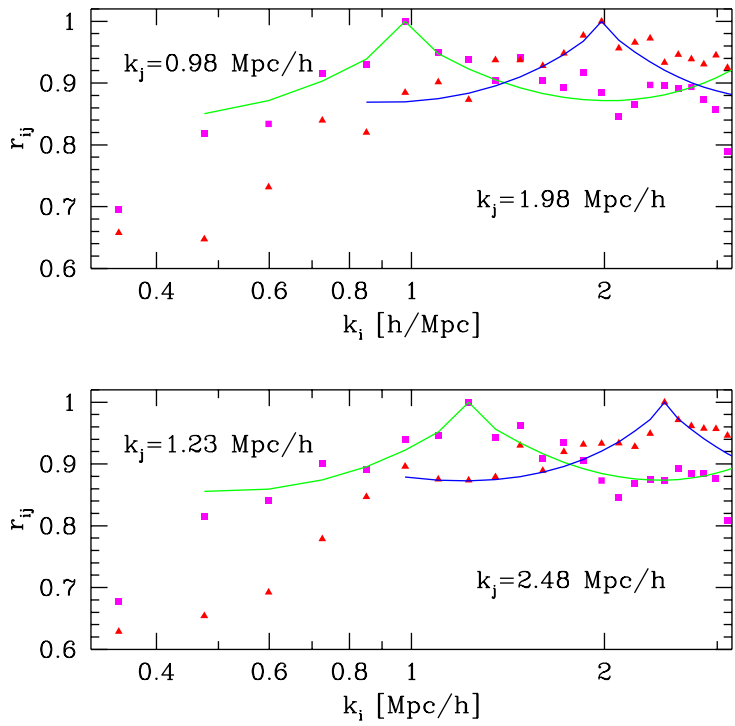}
\end{center}
\caption{Correlation coefficients $r_{ij}$ in the non-linear regime
predicted by HEPT (solid lines) for $R_a \approx R_b$ compared with
those from numerical simulations (symbols).  The width of shells in
$k$-space is $\d k=4\pi/100$ h/Mpc.}
\label{HEPT2}
\end{figure*}

\begin{figure*}
\begin{center}
\leavevmode
\epsfxsize=4.0in \epsfbox{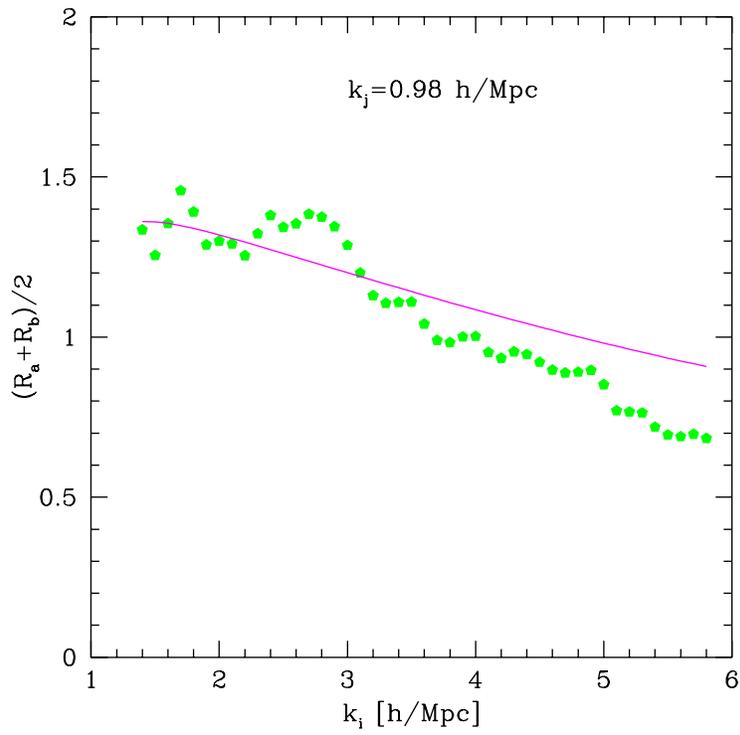}
\end{center}
\caption{Estimate of $(R_a+R_b)/2$ from the measured power-spectrum
covariance matrix, for $k_j=0.98$~h/Mpc, as a function of $k_i$, using
equations~(\protect\ref{HAdiag}-\protect\ref{HAndiag}).  The solid
lines show the expected $(R_a+R_b)/2$ value assuming that the
configuration dependence in the non-linear regime is the same as that
given by PT.}
\label{RaRb}
\end{figure*}

We have analyzed the covariance matrix for band-power estimates, both
for the three-dimensional matter density and its angular projection.
Three general statements can be made.  First, the non-linear nature of
gravitational clustering tends to increase the diagonal variance over
the Gaussian error as well as induce correlations between different
band-powers.  Second, for scales $k$ where $4 \pi k^3 P(k) \la 1$, the
covariance matrix is reasonably well approximated by its Gaussian
part; conversely, the Gaussian approximation rapidly breaks down on
scales $4 \pi k^3 P(k) \, \approxgt \, 1$.  Third, and interestingly,
band-powers on non-linear scales are actually significantly correlated
with band-powers on quasilinear scales (see e.g. figure \ref{PT2});
this is because the growth of wave modes on small scales is
significantly affected by the presence of long-wave modes.

We have also discussed how the relative importance of Gaussian and
non-Gaussian terms is somewhat dependent upon the choice of binning in
$k$-space, because the binning affects the Gaussian covariance but not
the non-Gaussian one (equation [\ref{cij}]). Coarse graining in
$k$-space lowers the Gaussian variance, while increasing the survey
size decreases the overall amplitude of the covariance matrix.
However, there are binning-independent ways to quantify the relative
importance of non-Gaussian versus Gaussian variances, such as by
focusing on the estimation of the amplitude of the power spectrum (see
\S \ref{parameter}). In this particular case, one is essentially using
a very coarse bin which includes all wave-modes. This is the reason
why simple estimates such as the one given in equation
(\ref{simpleest}) work as a rough guide as to the importance of
non-Gaussianity.  In the highly non-linear regime where all wave modes
are highly correlated, the amplitude of a rescaled power spectrum
(equation [\ref{hatydef}]) is more poorly determined compared to its
shape.  The key to decreasing error-bars in this case is to increase
the survey volume, rather than adding more wave-bands.

Angular projection adds interesting new twists to the above general
picture.  We have studied the projection relevant for weak lensing in
more detail, but most of our conclusions should hold for other
projections as well since the relevant weight function $w(\chi)$
(equation [\ref{proydef}]) drops out of the non-Gaussian-to-Gaussian
ratio $\cal R$ (equation [\ref{Rww}]).  Projection tends to reduce the
relative importance of non-Gaussianity, as fewer non-linear modes
contribute than in 3D.  In particular, for the cluster-normalized SCDM
model, even though the non-Gaussian terms eventually dominate the
covariance of the 3D power spectrum on small enough scales, they never
dominate for the 2D case, though they are comparable to the Gaussian
terms on certain angular scales. However, the amplitude of the effects
we study is very sensitive to the cosmological model, in particular to
the normalization $\sigma_8$ and the fluctuation growth rate.
Cluster-normalized low matter density models tend to show stronger
signs of non-Gaussianity, and open models even more so than $\Lambda$
models. We find that for most reasonable models the non-Gaussian terms
in the covariance grow significantly compared to the Gaussian terms
around $l\sim 1000$ which corresponds to $\theta\sim 3^{'}$.  They
even dominate for the open CDM model. This will have a significant
impact on the analysis of future weak-lensing surveys.

To get a handle on the size of the various effects we are interested
in, we have used perturbation theory for the linear or quasilinear
scales, and the hierarchical ansatz and numerical simulations for the
non-linear regime. As an outgrowth of this investigation, we have
shown that the hierarchical ansatz cannot be valid with $R_a$ and
$R_b$ strictly constant, for the particular type of trispectrum
configurations relevant for the power-spectrum covariance.  If this
simple model were correct, then the correlation coefficient $r_{ij}$
(equation [\ref{PT2}]) would become larger than one for two widely
separated shells, unless $R_a=-R_b$.  But it turns out that $R_a = -
R_b$ gives the wrong shape of $r_{ij}$ as a function of shell
separation.  Hence, the hierarchical ansatz cannot be valid in its
simplest form. We argue this failure is a consequence of the fact that
even in the non-linear regime the covariance matrix receives
contributions from large scales, and show how modifications of the
hierarchical ansatz based on PT can be made to give predictions that
match our numerical results.

There are at least two sets of issues we have left untouched here.
First, we have ignored the effect of biasing in our calculations.  A
simple linear biasing is of course trivial to include.  In this case
the relevant ratio of Gaussian to non-Gaussian terms in the covariance
is $V_s T_g/2 P_g^2 =V_s T/2 P^2\propto k^3 P=k^3 P_g/b^2$, assuming
the hierarchical ansatz, where $P_g$ and $T_g$ stand for the power
spectrum and trispectrum of the galaxies and $b$ is the linear bias
parameter.  Clearly the non-Gaussian effects we study are induced by
gravity, so the relevant quantity that governs their amplitude is the
size of the matter, not galaxy, fluctuations.  The value of $k$ where
the error-bars start to become larger than what is expected for a
Gaussian random field marks the non-linear scale.  Thus if $k^3P_g$ is
different from one at this scale we can conclude that there is a
significant bias between the mass and galaxy fluctuations, and in the
simple linear bias model we could in principle try to measure $b$
through this effect.  In reality, however, the biasing relation is
likely to be complicated by non-linearity and stochasticity,
especially on scales where the non-Gaussian covariance is not
negligible.

Lastly, we have focused throughout this paper on the sample-variance
dominated regime. In reality, shot-noise, either due to the discrete
nature of galaxies or to their random orientations, might be
non-negligible. Moreover the survey window will be important for
scales approaching the size of the survey.  It is relatively
straightforward to generalize our expressions to include these
effects. An interesting question is how to obtain the optimal
weighting of the data in the presence of non-Gaussianity.  This will
be presented in a separate paper.

\acknowledgements

We thank Josh Frieman, Andrew Hamilton, Wayne Hu, Jim Peebles, Uros
Seljak, Max Tegmark and Ludo van Waerbeke for useful discussions.
R.S. thanks the Institute for Advanced Study for hospitality, and
L.H. thanks the Canadian Institute for Theoretical Astrophysics, where
part of this work was done. While this work was nearing completion, we
became aware of work done by Meiksin and White (1998) who reached some
of the same conclusions.  We thank them for sending us their
preprint. In response to an earlier version of this paper, they now
include a similar calculation of the covariance matrix using the
hierarchical ansatz. Unfortunately the amplitudes were taken from {\em
spherically averaged} PT for {\em general} trispectrum configurations
(Fry 1984), rather than those relevant to the covariance matrix. The
disagreement they found with numerical simulations is a reflection of
these additional assumptions, as our results show. L.H. is supported
by the DOE and the NASA grant NAG 5-7092 at Fermilab.  M.Z. is
supported by NASA through Hubble Fellowship grant HF-01116.01-98A from
STScI, operated by AURA, Inc. under NASA contract NAS5-26555.

%
%
%=================
%\begin{references}

%\end{references}

\newpage

\begin{figure*}
\begin{center}
\leavevmode
\epsfxsize=4.0in \epsfbox{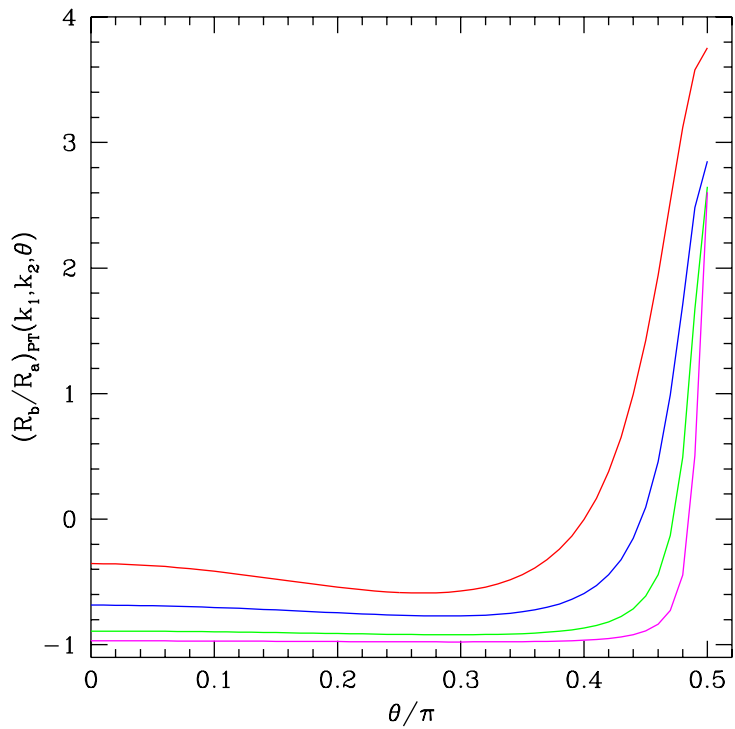}
\end{center}
\caption{The ratio $R_b/R_a$ given by PT as a function of
configuration angle $\theta$ between ${\k}_1$ and ${\k}_2$ for
different scales.  From top to bottom, $(k_1,k_2)=(1,2)$~h/Mpc,
$(1,4)$~h/Mpc, $(1,8)$~h/Mpc, and $(1,16)$~h/Mpc.  Note how the
relation $R_a=-R_b$ eventually holds for most configurations as the
shells become very distant.  }
\label{Rb/Ra}
\end{figure*}

\begin{figure*}
\begin{center}
\leavevmode
\epsfxsize=4.0in \epsfbox{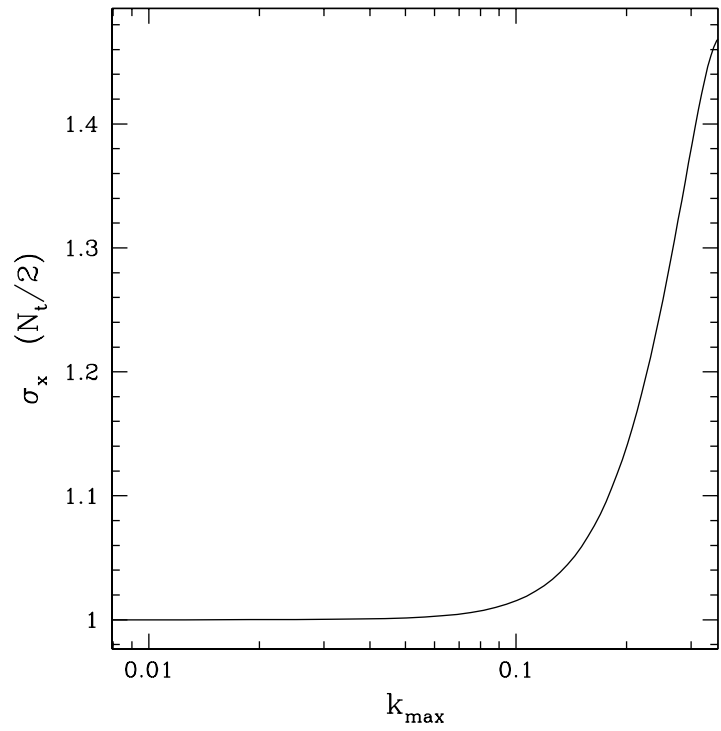}
\end{center}
\caption{Variance in $\hat x$ as a function of $k_{max}$.  Note that
the signature of non-Gaussianity is much larger here than what could
be seen in the diagonal terms coefficients shown in figure \ref{PT1}.
This is a result of the cross-correlations between bins.  }
\label{PT3}
\end{figure*}

\begin{figure*}
\begin{center}
\leavevmode
\epsfxsize=4.0in \epsfbox{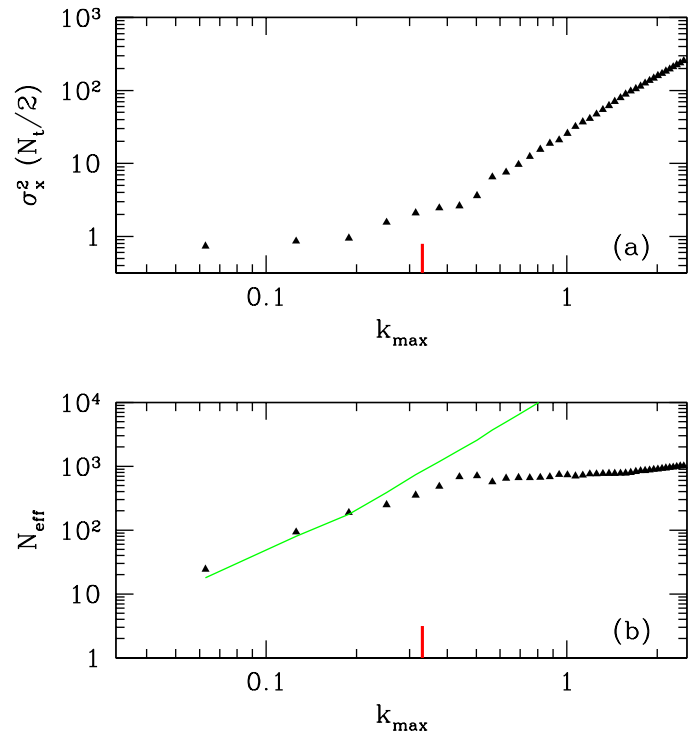}
\end{center}
\caption{Panel (a) shows the variance of $\hat x$ measured using the
N-body simulations.  Panel (b) shows the effective number of modes.
The results are for $\sigma_8=0.60$ ($z=0$) and the non-linear scale
is shown with a mark on the $x$ axis.}
\label{nb4}
\end{figure*}

\begin{figure*}
\begin{center}
\leavevmode
\epsfxsize=4.0in \epsfbox{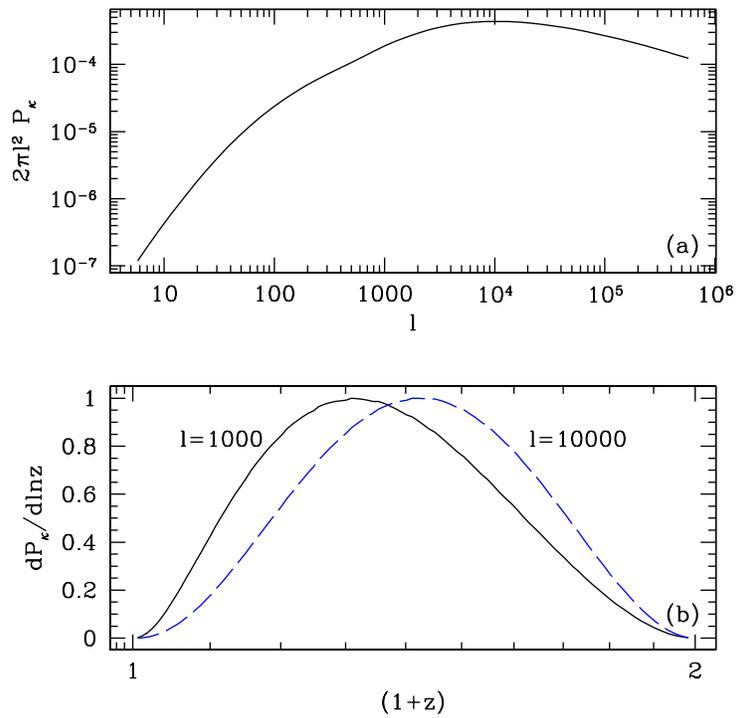}
\end{center}
\caption{Panel (a) shows the power spectrum of the projected mass
density for the cluster-normalized SCDM model. The sources are assumed
to be located at $z=1$. The bottom panel illustrates where the
contribution is coming from in redshift. The area under the curve is
proportional to the projected power spectrum, the normalization is
arbitrary.}
\label{lensing1}
\end{figure*}

\begin{figure*}
\begin{center}
\leavevmode
\epsfxsize=4.0in \epsfbox{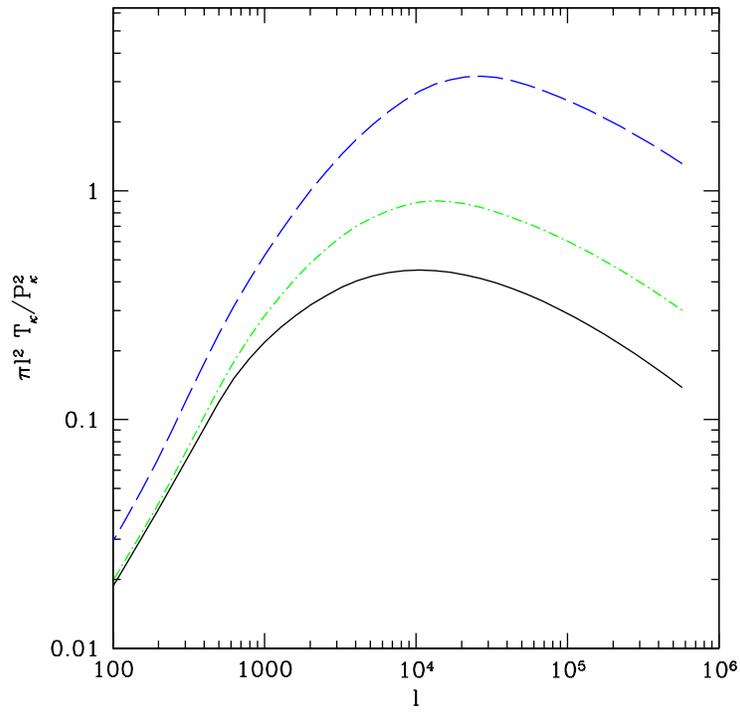}
\end{center}
\caption{Ratio of the non-Gaussian to Gaussian terms in the covariance
of the projected mass density, for cluster-normalized SCDM (solid
line), $\Lambda$CDM (dot-dashed line) and open CDM (dashed line)
models. The hierarchical ansatz was assumed in the calculation, with
$R_a \approx R_b$. See text for details.  }
\label{lensing2}
\end{figure*}

\end{document}